\documentclass[a4paper,11pt,final]{article}

\usepackage{authblk}

\usepackage{graphics}
\usepackage{graphicx}
\usepackage{epsfig}

\usepackage{amsthm}

\usepackage{amsmath}
\usepackage{amssymb}

\date{}

\begin{document}

\title {Generalized Poisson-Kac processes and hydrodynamic modeling
of systems of interacting particles I - Theory}

\author{Massimiliano Giona}
\affil{Dipartimento di Ingegneria Chimica DICMA
Facolt\`{a} di Ingegneria, La Sapienza Universit\`{a} di Roma
via Eudossiana 18, 00184, Roma, Italy  \authorcr
Email: massimiliano.giona@uniroma1.it}

\maketitle

\begin{abstract}
This article analyzes the formulation
of space-time continuous hyperbolic hydrodynamic
models for systems of interacting particles moving on a
lattice, by connecting their local stochastic lattice
dynamics to the formulation of an associated (space-time continuous)
Generalized Poisson-Kac process possessing the same local
transition rules. The hyperbolic hydrodynamic limit
follows naturally from  the statistical description of
the latter in terms of the system of its partial probability
density functions.
Several cases are treated, with particular
attention to: (i) models of interacting particles
satisfying an exclusion principle, and (ii)  models
defined by a given interparticle interaction potential.
In both cases, the hydrodynamic models may display
singularities, dynamic phase-transitions and bifurcations
(as regards the flux/concentration-gradient constitutive
equations), whenever the Kac limit of the model (infinite
propagation velocity limit) is considered.
\end{abstract}

\section{Introduction}
\label{sec1}

The study of systems of interacting particles represents
a central issue in the thermodynamics of irreversible processes
and in transport theory since the seminal work by Boltzmann
on the kinetic theory of dilute gases \cite{boltzmann,balescu}.

In many cases the analysis of this problem
can be simplified by considering 
 particle motion on a discrete lattice.
In this way, local particle dynamics is expressed as a system
of transition probabilities for particle hopping
between the nearest neighburing sites of the lattice. 
For the setting of this class of problems the reader is
referred to  \cite{or1,or2,or3,or4}.

In lattice problems, interactions depend either on sterical
and quantum effects, or by the explicit representation
of the interaction potential. Sterical
and quantum effect imply some form of exclusion
principle, whenever no more that a single particle
or at most  a finite number of particles with different
values of the some internal degree of freedom (spin)
can be simultaneously present at the same lattice site.
Interaction potentials, be them short or long-ranged, influences the
hopping transition matrix in a continuous way.

One of the central issues in the physical understanding
of these particle systems is the description of their
collective statistical properties, i.e., the transition
from the local probabilistic lattice dynamics at the level
of the single lattice site to a continuous space-time evolution
for the associated concentration field (probability density
function), accounting for the collective motion
of a statistical particle ensemble.

The transition from the lattice motion to the continuous
and collective description of particle dynamics, which is 
the key problem in statistical physics, involves essentially two
different and conceptually separate steps:
(i) the collective description of the interaction amongst particles
in the form of  constitutive equations for the probability
density flux, expressed as generic nonlinear
functional of the particle concentration field, of its spatial
gradient and, in principle  of its spatial derivatives of any
order, and (ii) the continuum limit of a lattice particle problem,
the time evolution of which is defined at discrete  time
instants, in the form of a physical system defined in a continuous
space-time.

The first problem is in general extremely difficult and its
solution often requires suitable  physical approximations
on the  representation of particle interactions
in terms of functionals involving the particle probability
density function (one-particle density).
The classical example of this type of approximation is the {\em stosszahlansatz}
in the Boltzmannian description, in which the effects
of the binary collisions are treated (invoking the hypothesis
of molecular chaos) as loss and gain terms in the
evolution equation for the one-particle distribution
function and  can be assumed proportional to the product of the two one-particle
distribution functions $f({\bf q},{\bf v},t)$, $f({\bf q},{\bf v}^\prime,t)$
performing a collisional event with velocities ${\bf v}$ and 
${\bf v}^\prime$. 
A similar approximation characterizes
the kinetic theory of other systems such e.g.
a gas of electrons (plasma), where a self-consistent
continuous approximation for the electric field
is adopted in the Vlasov equation \cite{balescu}.

An example of the latter problem (transition from a lattice
to a continuum description) is the statistical
formulation in a spece-time continuum of
lattice random walk for system of independent particles,
i.e., in the absence of exclusion principles or potential
contributions \cite{giona_lrw}. The technical issue
in this case in the transformation of the discrete Markov process
describing the evolution for the probability
density function of particles evolving onto the lattice
(characterized by a discrete spacing $\delta$ between
nearest neighbouring sites) at discrete times 
(corresponding to a  physical time interval $\tau$
between subsequent events), into a continuous group (or semigroup)
of tranformations parametrized with respect to the physical time 
$t \in {\mathbb R}^+$ acting on the  probability density functions
$p(x,t)$, continuously parametrized with respect to
the space coordinate $x \in {\mathbb R}$.

The latter problem involves the so called {\em hydrodynamic
limit}, defined for lattice spacing $\delta$ and characteristic
time $\tau$ tending to zero, assuming a suitable
scaling ansatz between the two characteristic space-time
parameters, expressed in the form of a limit
behavior
\begin{equation}
\lim_{\tau \rightarrow 0} \frac{\delta^{\alpha}(\tau)}{\tau}
= \mbox{constant}
\label{eq1_1}
\end{equation}
where $\alpha>0$ is some characteristic exponent
defining the scaling ansatz.
For a thorough discussion on the mathematical physical
aspects  of the setting and formulation
of the hydrodynamic limit for lattice particle
dynamics  and on the functional form of the
resulting hydrodynamic models for prototypical
interacting particle systems, the reader is referred
to the classical monographs on this topic \cite{landim,presutti}.

In principle, different choices of the scaling assumption (\ref{eq1_1})
provides different hydrodynamic models as analyzed in \cite{giona_lrw},
and briefly reviewed in Section \ref{sec2}.
Some choices of the scaling ansatz (\ref{eq1_1}), and
specifically the diffusive scaling corresponding to $\alpha=2$
destroy some fundamental physical properties associated
with lattice propagation, and forces the hydrodynamic
formulation of the statistical properties
of the system to be described by parabolic
models (first-order in time, second-order in space derivatives)
that, by nature, violates fundamental physical conditions
(finite propagation velocity, deriving
from the Minkowskian metrics of the space-time).

The latter hydrodynamic approach (leading to parabolic models)
 is fully rigorous from
the mathematical point of view. Nevertheless,
it superimposes and intermingles two qualitative different
physical properties: (i) the existence of long-term (emerging)
statistical features in a lattice particle systems, with (ii)
the formulation of a  continuous space-time  description
of its statistical evolution, defined technically from
the operation of  letting $\delta, \,\tau \rightarrow 0$
with the constraint imposed by the scaling assumption.

From the physical point of view, the assessment of a continuous
limit  is in principle independent of the finite/infinitesimal
 values of $\delta$ and $\tau$. More precisely,
there are situations, in which the lattice description
is an approximation of the continuous evolution of
a particle system in which the values of the
parameters $\delta$ and $\tau$  do possess a well
defined physical meaning, and are not allowed to
attain vanishing values. A typical situation
of this sort is a diluted particle gas system,
where, near equilibrium,
 $\delta$ corresponds to the mean-free path $\lambda(T,P) \sim T/P$
between two subsequent collisions depending on the
temperature $T$ and on the pressure $P$, while the
characteristic lattice time scale $\tau$ is related
to the root mean square speed $v_{\rm rms}(T) \sim T^{1/2}$ 
depending solely on temperature.
A diffusive scaling ansatz ($\alpha=2$) would implies
$T^{3/2}/P=\mbox{constant}$, which violates the
equilibrium gas law in diluted condition $P/T=\mbox{constant}$
for fixed volume and particle number.

The analysis developed in \cite{giona_lrw} for the
random walk of independent particles on a lattice suggests
another possibility for deriving a space-time continuous
statistical description of a system of particles on a lattice
for any finite value of $\delta$ and $\tau$, respectful
of the local lattice dynamics. The tool for achieving this program,
at least for lattice dynamics of independent particles,
is the connection of the original lattice equation of motion
with an associated Generalized Poisson-Kac process 
possessing the same transition probabilities amongst
local directions of motion, out which a space-time continuous
statistical description of the original lattice process follows.

The scope of the present work is to develop a similar
program for systems of interacting particles, which is
a much more challenging task as the local dynamic rules for particle
motion depend on the state of the whole particle ensemble.
These collective effects can be formally treated by invoking a molecular
chaos assumption similar to the Boltzmannian ``stosszahlansatz''
(see Section \ref{sec4}).

Once the statistical  description of systems of interacting particles
has been embedded in the theory of GPK processes
new physical phenomenologies can be unveiled, associated
with: (i) the Kac limit of the resulting hyperbolic
hydrodynamic description whenever the characteristic
propagation velocity is hypothesized to diverge (this
occurs for particle systems subjected to exclusion principles);
(ii) a new class of dynamic phase transitions can occur
in the presence of interparticle potentials, related
to multiplicity and bifurcations in the constitutive
equations for the concentration flux in terms of the
concentration gradient.

Throughout this article the theory is developed
for system of interacting particles in one-dimensional
spatial problems, in order to simplify the
notation and highlight in the simple possible
way the new and rich phenomenology that can occur.
The numerical investigation of the main qualitative phenomenologies
highlighted in this article is addressed in \cite{giona_numerics},

The article is organized as follows. Starting from a brief
conceptual summary of the result presented in \cite{giona_lrw},
section \ref{sec2} reviews the formalism of Generalized Poisson-Kac
processes, and its application to achieve a hyperbolic
continuous statistical description of interacting particle
systems. Section \ref{sec3} analyzes the
construction of the corresponding GPK processes
for systems of particles satisfying an exclusion
principles. The analysis is limited to the
case of a tagged particle in a mean field characterized
by a given (and fixed) particle concentration.
Section  \ref{sec4} extends the analysis
to the nonlinear case. The class of models considered
corresponds to exclusion models where
the exclusion principle is satisfied probabilistically.
This concept is introduced in this Section and thoroughly explained.
The resulting nonlinear hyperbolic hydrodynamic models
display very interesting and singular features in
the Kac limit. 
Finally section \ref{sec5}
develops the formalism of hyperbolic hydrodynamic models
in the presence of interaction potentials.

\section{Stochastic processes with finite propagation velocity and hydrodynamic behavior}
\label{sec2}

In a recent work \cite{giona_lrw}, Giona analyzed a very
simple example of lattice particle dynamics: the random
walk of independent  particles on a one-dimensional lattice
in the case of asymmetric transitions amongst the two nearest
neighboring sites (Asymmetric Lattice Random Walk, ALRW)
and its continuous statistical description.
The discrete lattice dynamics is characterized by
the lattice spacing $\delta$ between nearest neighboring
sites  and by the constant hopping time $\tau$ between
two subsequent events.

The starting observations motivating this revisitation
of ALRW are:
\begin{itemize}
\item the definition of a space-time continuous process associated
with ALRW does not require the limit for $\delta$ and $\tau$
tending to zero. This is because a time-continuous
formulation of the process  requires solely the local
interpolation of particle trajectories between subsequent time instants
$t_n= n \, \tau$, and $t_{n+1}=(n+1) \, \tau$ and subsequent
positions $x_{n}$, $x_{n+1}$, and eventually the assumption
of some level of uncertainty in the initial particle position
$x_{0}$.
\item The long-term emergent statistical properties
of the process are well defined for any (finite and non vanishing)
values of $\delta$ and $\tau$. Consequently, a space-time
continuous hydrodynamic model for this process should be defined
independently of any lattice limit $\delta, \, \tau \rightarrow 0$,
and of any scaling ansatz connecting $\delta$ and $\tau$
in this limit.
\item In a smooth, time-continuous, formulation of the
process, the ratio  $b_0=\delta/\tau$, corresponding to the
local propagation velocity, should be constant and bounded.
\item A time-continuous hydrodynamic model, subjected to
the above mentioned constraint on the local
propagation velocity, should be able to describe the whole
process dynamics, from the early stages, at which 
particles perform a ballistic motion, to
the long-term dispersive features, corresponding to
a linear Einsteinian scaling of the mean square displacement,
for any value of $\delta$ and $\tau$.
\end{itemize}
It has been shown in \cite{giona_lrw} that the formulation 
of such a ``smooth'' hydrodynamic model is possible
and it is grounded on the formulation
of a space-time continuous stochastic process, analogous to
ALRW, belonging to the class of Generalized Poisson-Kac processes
\cite{gpk0,gpk1,gpk2,gpk3}.
Here the diction ``smooth'' has been used to indicate
that the local propagation velocity is bounded, contrarily to
the classical limit formulation grounded on a diffusive
scaling asumption $\delta^2/\tau=\mbox{constant}$, leading
to a stochastic description based on almost nowhere differentiable
Wiener processes.
In the next paragraph, the basic concept of GPK theory are reviewed.

\subsection{Generalized Poisson-Kac processes}
\label{sec2_1}
The introduction of Generalized Poisson-Kac processes (GPK for short) stems
originally from two main physical reasons:
(i) to generalize the class of stochastic
models proposed by Marc Kac in one-dimensional spatial
systems \cite{kac}, possessing finite propagation velocity
and driven by a simple Poisson process, to any spatial
dimension and to any number of stochastic states (including the
limit towards a continuum of states); (ii) the setting of stochastically
consistent transport models of hyperbolic nature suitable
for describing physical transport processes possessing
finite propagation velocity.
Here, ``stochastically consistent'' means that there exists
a stochastic process admitting these models as its statistical
description. This issue is closely   connected to the
fact, that while the original one-dimensional model
considered by Kac provides a stochastic interpretation
for the one-dimensional Cattaneo equation  
$\partial_t p(x,t) + \tau_c \, \partial_t^2 p(x,t) = D \,  \partial_x^2 p(x,t)$, where $\tau_c$ and $D$ are 
positive constants and $\partial^\alpha_\xi=\partial^\alpha/\partial \xi^\alpha$,
$\xi=t,x$, $\alpha=1,2$ \cite{cattaneo},
there are no stochastic processes in ${\mathbb R}^n$ with $n \geq 2$
admitting the higher dimensional Cattaneo
model $\partial_t p({\bf x},t) + \tau_c \, \partial_t^2 p({\bf x},t) = D
\,  \nabla^2 p({\bf x},t)$ as the evolution equation
for their probability density function $p({\bf x},t)$.
This property follows also from the observation that the
Green function for the Cattaneo hyperbolic transport model
in ${\mathbb R}^n$, $n \geq 2$ does not present positivity and
attains negative values \cite{cattaneo_criticism}
 (which is deprecable in a probabilistic
context). The definition of GPK processes is closely
connected with the class of higher-dimensional stochastic
models studied by Kolesnik \cite{kolesnik1,kolesnik2,kolesnik3}

A GPK process in ${\mathbb R}^n$
is defined by a finite number $N$ of stochastic states,
by a family of $N$ constant velocity vectors $\{ {\bf b}_h \}_{h=1}^N$,
${\bf b}_h \in {\mathbb R}^n$, by a vector of transition
rates ${\boldsymbol \Lambda}=(\lambda_1,\dots,\lambda_N)$, $\lambda_h>0$.
$h=1,\dots,N$, and  by a $N \times N$ transition probability matrix
${\bf A}=(A_{h,k})_{h,k=1}^N$, $A_{h,k} \geq 0$, $\sum_{h=1}^N A_{h,k}
=1$, $\forall k=1,\dots,N$.
The generator of stochasticity is a finite $N$-state Poisson
process $\chi_N(t;{\boldsymbol \Lambda},{\bf A})$ attaining $N$
distinct values $\chi_N=1,,\dots,N$, and such that
the probabilities $\widehat{P}_h(t)= \mbox{Prob}[\chi_N(t)=h]$,
$h=1,\dots,N$ satisfy the Markov chain dynamics
\begin{equation}
\frac{d \widehat{P}_h(t)}{d t} = - \lambda_h \, \widehat{P}_h(t)
+ \sum_{k=1}^N A_{h,k} \, \lambda_k \, \widehat{P}_k(t)
\label{eq2_1}
\end{equation}  
From the above setting it follows that a GPK process ${\bf X}(t)$
in ${\mathbb R}^n$ is 
defined by the stochastic differential equation
\begin{equation}
d {\bf x}(t)= b_{\chi_N(t;{\boldsymbol \Lambda},{\bf A})} \, dt
\label{eq2_2}
\end{equation}
This means that according to the transition mechanism of state
recombination specified by the $N$-state finite Poisson process
$\chi_N(t;{\boldsymbol \Lambda},{\bf A})$ , defined
by ${\boldsymbol \Lambda}$ and ${\bf A}$, the velocity vector
defining eq. (\ref{eq2_1}) switches amongst the $N$
possible realizations ${\bf b}_1,\dots,{\bf b}_N$.

Since $\max_{h=1,\dots,N} |{\bf b}_h| \leq B$
is bounded, the process possesses finite propagation velocity
and  the trajectory ${\bf x}(t)$ 
of each realization of a GPK process is with 
probability 1 an almost everywhere smooth function of time
consisting of smooth line segments. It is therefore
differentiable at all the time instant, but at
the transition points, where  $\chi_N(t;{\boldsymbol \Lambda},{\bf A})$ 
switches from one state to another, still possessing
well defined left and right derivatives at the transition points
(Lipschitz continuity).

The statistical description of a GPK process involves $N$
partial probability density functions $p_h({\bf x},t)$,
$h=1,\dots,N$,
\begin{equation}
p_h({\bf x}, t) \, d {\bf x} =
\mbox{Prob} \left [ {\bf X}(t) \in ({\bf x},{\bf x}+d {\bf x}),
\; \; \chi_N(t)=h \right ]
\label{eq2_3}
\end{equation}
where ${\bf x}=(x_1,\dots,x_N)$, 
${\bf X}(t)=(X_1(t),\dots,X_N(t))$,
$d {\bf x}= \prod_{h=1}^N d x_h$
is the measure element, and ${\bf X}(t) \in ({\bf x},{\bf x}+d {\bf x})$
means that for each $X_h(t)$,  $X_h(t) \in (x_h,x_h+dx_h)$, 
$h=1,\dots,N$. The partial probability densities satisfy the
system of first-order  differential equations
\begin{equation}
\frac{\partial p_h({\bf x},t)}{\partial t}
= - {\bf b}_h \cdot \nabla p_h({\bf x},t)
-\lambda_h \, p_h({\bf x},t)
+ \sum_{k=1}^N A_{h,k} \, \lambda_k \, p_k({\bf x},t)
\label{eq2_4}
\end{equation}
Eq. (\ref{eq2_4}) represents the complete statistical description
of a GPK process: it plays the same role of the classical
parabolic  Fokker-Planck
equation for  Langevin  models driven by Wiener noise.
The difference with the latter case is that, for GPK processes,
 a system
of $N$ partial probability densities, accounting also
for the local state of the stochastic perturbation should be
defined, owing to the non strictly Markovian structure
of the process. The overall probability density function of the process
is $p({\bf x},t)=\sum_{h=1}^N p_h({\bf x},t)$, and satisfies
the conservation equation
\begin{equation}
\frac{\partial p({\bf x},t)}{\partial t}
= -\nabla \cdot {\bf J}_p({\bf x},t)
\label{eq2_5}
\end{equation}
where the probability density flux ${\bf J}_p({\bf x},t)$
is expressed by
\begin{equation}
{\bf J}_p({\bf x},t)= \sum_{h=1}^N {\bf b}_h \, p_h({\bf x},t)
\label{eq2_6}
\end{equation}
and the constitutive equation for ${\bf J}_p({\bf x},t)$, follows from
the definition (\ref{eq2_6})  and from the balance equations (\ref{eq2_4}).

Depending on the structural properties of the GPK,
i.e., on $\{ {\bf b}_h\}_{h=1}^N$ ${\boldsymbol \Lambda}$ and
${\bf A}$, a variety of different stochastic models
can be constructed and the reader is  referred to \cite{gpk1} for
a structural characterization of these processes.
Consider below the simple case where all the stochastic velocity
vectors possess the same modulus $b_0$, i.e.,
${\bf b}_h = b_0 \, {\bf e}_h$, $h=1,\dots,N$,
 where ${\bf e}_h$ are unit vectors and all the
transition rates are equal i.e., $\lambda_h= \lambda_0$,
$h=1,\dots,N$. Under this conditions, it is natural to
formulate the {\em Kac limit} of a GPK process, i.e.,
the asymptotics of the GPK process in the case $b_0, \,
\lambda_0 \rightarrow \infty$, keeping fixed the
ratio
\begin{equation}
\frac{b_0^2}{2 \, \lambda_0} = D_{\rm nom}
\label{eq2_7}
\end{equation}
where $D_{\rm nom}$ is referred to as the ``nominal diffusivity''
of the GPK process.
The Kac limit corresponds to the limit behavior of a GPK
process in the case its propagation velocity
diverges and the same does the transition rate, under
the scaling hypothesis (\ref{eq2_7}).
Under this conditions, and assuming reasonable no-bias
constraints on the system of velocity vectors $\{{\bf e}_h\}_{h=1}^N$
(see \cite{gpk1} for details), the balance equations (\ref{eq2_4})
for the partial probability density functions $p_h({\bf x},t)$
collapse into a single  parabolic equation for the
overall probability density $p({\bf x},t)$
\begin{equation}
\frac{\partial p({\bf x},t)}{\partial t}= \nabla \cdot \left 
( {\bf D} \, \nabla p({\bf x},t) \right )
\label{eq2_8}
\end{equation}
where ${\bf D}=(D_{h,k})_{h,k=1}^N$ is the
effective diffusivity tensor. If the system possesses
enough symmetries, ${\bf D}$ is isotropic, i.e.,
${\bf D}= D_0 \, {\bf I}$, and the Kac limit of the process
is characterized by the single overall probability
density function $p({\bf x},t)$, solution of the
diffusion equation
\begin{equation}
\frac{\partial p({\bf x},t)}{\partial t}= D_0 \, \nabla^2 p({\bf x},t)
\label{eq2_9}
\end{equation}
where $D_0$ is the scalar effective diffusivity,
depending linearly on $D_{\rm nom}$, i.e.
$D_0= D_{\rm nom} \, \kappa$, where $\kappa \sim  {\mathcal O}(1)$.

In the case of ALRW ($n=1$),  the number of states is $N=2$,
corresponding to the movements towards the two (left
and right) neighboring sites of any lattice site. Correspondingly,
the velocity vectors are expressed by $b_1=b_0$,
$b_2=-b_0$, where $b_0=\delta/\tau$.
As regards the transition probabilities, if $r_1$ and
$r_2$   are  probabilities of moving to the right/left site respectively,
 letting $r=r_1-r_2$, it follows that the transition
probability matrix  is
given by
\begin{equation}
{\bf A}=
\left (
\begin{array}{cc}
\frac{1+r}{2} & \frac{1+r}{2} \\
\frac{1-r}{2} & \frac{1-r}{2}
\end{array}
\right )
\label{eq2_10}
\end{equation}
The GPK process associated with the ALRW dynamics on the
real line is  thus expressed by
\begin{equation}
d x(t)= b_{\chi_2(t; \lambda_0  \, {\bf I}, {\bf A})} \, d t
\label{eq2_11}
\end{equation}
where the transition rate vector ${\boldsymbol \Lambda}= \lambda_0 \, {\bf I}$
is isotropic and characterized by the value $\lambda_0$.
The expression for $\lambda_0$ in terms of the lattice
parameters can be obtained from the
long-term linear scaling of the mean square displacement
in the simplest case of symmetric motion
($r=0$) for which $\lambda_0=2/\tau$ follows.

The system of hyperbolic
first-order equation for the partial probability
densities $p_1(x,t)$, $p_2(x,t)$ represents
a continuous hydrodynamic model for the statistical
properties of ALRW, and the classical hydrodynamic limit
(see e.g. \cite{weiss}) can be regarded as
the Kac limit of this hyperbolic model.

The latter observation provides a  novel way
of  interpreting the classical parabolic hydrodynamic limit
of lattice particle dynamics: not as the limit
for space-time discretized characteristic scales ($\delta$ and
$\tau$) tending to zero (as  eq. (\ref{eq2_11}) is
already defined in a space-time continuum $(x,t) \in {\mathbb R} \times {\mathbb R}^+$), but as the limit for the characteristic propagation
velocity of the process $b_0$ tending to infinity, assuming
also that the transition rate would diverge $\lambda_0 \rightarrow \infty$.
In the latter (Kac) limit, the scaling relation (\ref{eq2_7})
is essential in ensuring the existence of this limit.
For further details see \cite{giona_lrw}.

\subsection{The program}
\label{sec2_2}

\begin{figure}[!]
\begin{center}
\epsfxsize=10.cm
\epsffile{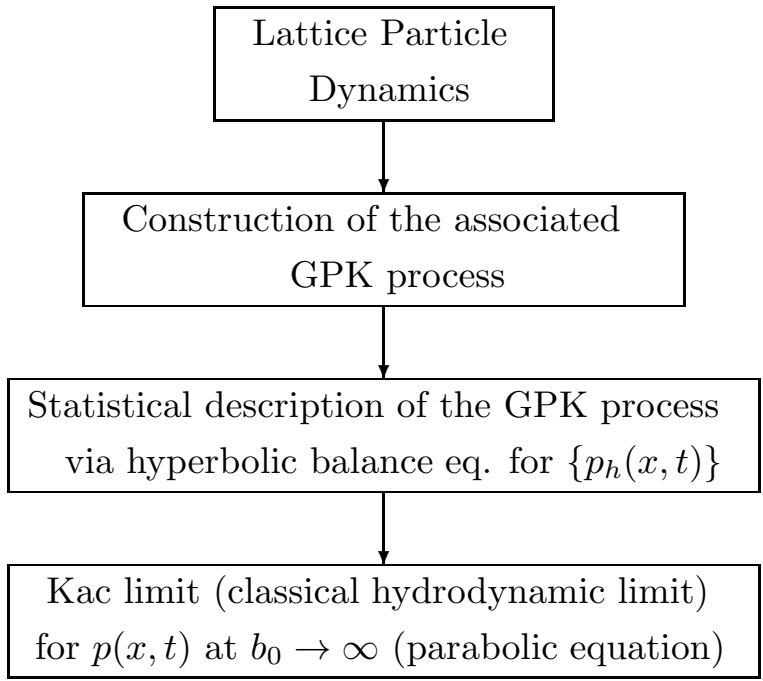}
\end{center}
\caption{Program towards the hyperbolic formulation
of continuous hydrodynamic models of interacting particle systems.}
\label{Fig0}
\end{figure}

From the analysis developed above, it follows
a conceptual program towards the construction
of continuous hydrodynamic models of systems of interacting particles.
This program is reviewed schematically in figure \ref{Fig0},
and follows the same approach applied in \cite{giona_lrw} to ALRW.

The central issue is the association with a local
lattice dynamics of its corresponding continuous GPK process, possessing the
same transition probability structure of the original lattice model.
Once this step is performed, the derivation of the different
forms of continuous hydrodynamic models follows directly from GPK
theory. In the remainder of this article, this program 
is outlined and developed for prototypical models 
of interacting particle systems.

\section{Model systems and mean field analysis of tagged particle diffusion}
\label{sec3}

In this Section we consider typical random walk models with exclusion,
meaning that at each lattice site no more than one particle or
a finite number of them, possessing different characteristic
properties (spin), can be present simultaneously.

For several prototypical models  we first derive
the mean-field behavior of a tagged particle i.e., the
properties of the particle diffusive motion by assuming that
the average particle concentration is given. Subsequently,
we provide the formalization of the same process within
the GPK formalism.

Throughout this Section, we consider one-dimensional spatial
models. 

\subsection{Fermionic random walk with exclusion}
\label{sec3_1}

This model has been addressed by Colangeli et al.  \cite{latent_heat}
and represents,
in the absence of other interactions, a form of Kawasaki model \cite{kawasaki}.
Particles behave as fermions, and the direction of the
velocity $\pm 1$, corresponds to their spin.
At each lattice site, at most two particles can be simultaneously 
present with opposite spins (i.e., oppositive velocity directions).

The dynamic of the exclusion interaction is as follows:
\begin{enumerate}
\item first, a velocity switch is considered, meaning that if solely a particle is present at a given site it switches its direction
with probability $1/2$;
\item the next step is the advective step: particles
at a given site move towards the nearest neighboring sites
consistently with their velocity directions, i.e., with
the values of their spins, and compatibly with the
exclusion principle. For instance, a particle at site $k$  possessing velocity $+1$ moves towards $k+1$ provided that the arrival site
does not  contain already  a particle with positive velocity.
\end{enumerate}

As stated at the beginning of this Section, consider the
self-diffusion dynamics of a tagged particle, assuming that the
average fraction of positively and negatively oriented
particles is  equal to  $\pi \in [0,1]$.

The random walk model of a tagged particle following the
recipe stated above, (which is a mean-field approximation),
can be described by considering at time
$n$ both the particle position $x_n$ and its spin variable $s_n$.

The dynamics for the spin variable is given by:
\begin{equation}
s_{n+1} = \xi_{n+1} \, s_n
\label{eq3_1}
\end{equation}
starting at time $n=0$ from $s_0=\{0,1\}$, where $\xi_n$
are uncorrelated random variables attaining
values $\pm 1$, according to the probabilistic
scheme
\begin{equation}
\xi_{n+1} =
\left \{
\begin{array}{lll}
1 & & \mbox{Prob} \; \, (1+\pi)/2 \\
-1 & & \mbox{Prob} \; \, (1-\pi)/2
\end{array}
\right .
\label{eq3_2}
\end{equation} 
For instance $\mbox{Prob}[\xi_{n+1}=-1]=(1-\pi)/2$, corresponding
to the probability of a velocity switching, equals the
probability that a switching event  occurs (which is
$1/2$) times the probability that the arrival site does not
contain already a particle with oppositive velocity
(which equals $1-\pi$).
Observe that the random variables $\xi_n$, $n=1,2,\dots$
are uncorrelated with each other, i.e.,
\begin{equation}
\langle \xi_h \, \xi_k \rangle =
\left \{
\begin{array}{lll}
\langle \xi_h^2 \rangle & &  k=h \\
\langle \xi_h \rangle \, \langle \xi_k \rangle & & 
k \neq h
\end{array}
\right .
\label{eq3_3}
\end{equation}
As regards the initial condition $s_0$, one has
\begin{equation}
s_0 =
\left \{
\begin{array}{lll}
1 & & \mbox{Prob} \; \, 1/2 \\
-1 & & \mbox{Prob} \; \, 1/2 
\end{array}
\right .
\label{eq3_4}
\end{equation}
so that $\langle s_0 \rangle =0$.

\begin{figure}[!]
\begin{center}
\epsfxsize=10.cm
\epsffile{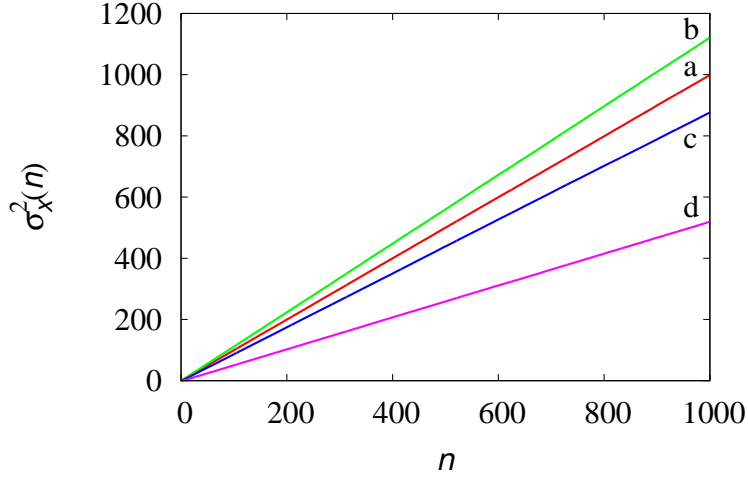}
\end{center}
\caption{Mean square displacement $\sigma_x^2(n)$ vs $n$ for
the fermionic random walk model with exclusion described in the
main text, deriving from the stochastic simulation
of eqs. (\ref{eq3_1}) and (\ref{eq3_5}). Line (a) refers to
$\pi=0$, line (b) to $\pi=0.2$, line (c) to $\pi=0.6$, line (d) to $\pi=0.8$.}
\label{Fig1}
\end{figure}

\begin{figure}[!]
\begin{center}
\epsfxsize=10.cm
\epsffile{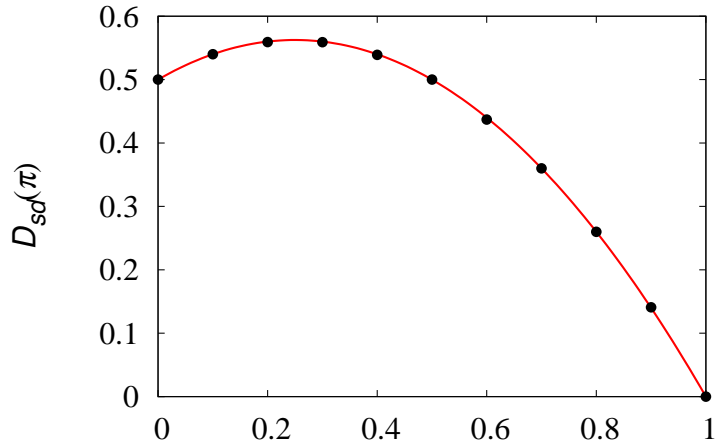}
\end{center}
\caption{Self-diffusion coefficient for the fermionic random walk
model with exclusion. Solid line corresponds to the
graph of eq. (\ref{eq3_17}), symbols ($\bullet$) refer to the
results of random walk simulations depicted in figure \ref{Fig1}.}
\label{Fig2}
\end{figure}

The dynamics of particle position  $x_{n}$ is then  expressed by
\begin{equation}
x_{n+1}=x_n+ s_{n+1} \, \eta_{n+1}
\label{eq3_5}
\end{equation}
where $\eta_{n+1}$ are random variables
attaining values $0,1$ according to the rule
\begin{equation}
\eta_{n+1} =
\left \{
\begin{array}{lll}
0 & & \mbox{Prob} \; \, \pi \\
1 & & \mbox{Prob} \; \, (1-\pi)
\end{array}
\right .
\label{eq3_6}
\end{equation}
For instance, $\mbox{Prob}[\eta_{n+1}=0]$ correponds to the probability that
the arrival site contains already a particle with the same
spin, and therefore equals $\pi$.
Also the variables $\eta_h$ are uncorrelated with
each other,
\begin{equation}
\langle \eta_h \, \eta_k \rangle =
\left \{
\begin{array}{lll}
\langle \eta_h^2 \rangle & &  k=h \\
\langle \eta_h \rangle \, \langle \eta_k \rangle & & 
k \neq h
\end{array}
\right .
\label{eq3_7}
\end{equation}
and independent of the $\xi_k$-variables, $\langle \eta_h \, \xi_k \rangle =
\langle \eta_h \rangle \, \langle \xi_k \rangle$.
Setting $x_0=0$, it follows from eqs. (\ref{eq3_1}) and (\ref{eq3_5}) that
\begin{equation}
x_n= \sum_{h=1}^n s_h \, \eta_h
\label{eq3_8}
\end{equation}
for $n \geq 1$, where
\begin{equation}
s_h = s_0 \, \prod_{k=1}^h \xi_k
\label{eq3_9}
\end{equation}
It follows that
\begin{equation}
\langle s_n \rangle =0 \, , \qquad
\langle s_h \, s_k \rangle = \langle s_0 \, s_{|k-h|} \rangle
\label{eq3_10}
\end{equation}
and therefore
\begin{equation}
\langle x_n \rangle =0 \, , \qquad
\langle x_n^2 \rangle = \sum_{h=1}^n \sum_{k=1}^n
\langle s_h \, s_k \rangle \, \langle \eta_h \, \eta_k \rangle
\label{eq3_11}
\end{equation}
Since
\begin{equation}
\langle \eta_h \, \eta_k \rangle =
\left \{
\begin{array}{lll}
1-\pi & & k=h \\
(1-\pi)^2 & & k \neq h
\end{array}
\right .
\label{eq3_12}
\end{equation}
and 
\begin{eqnarray}
\langle s_0^2 \rangle & = & 1 \nonumber \\
\langle s_0 \, s_m  \rangle &= & \prod_{k=1}^m \langle \xi_k \rangle = \pi^m
\; , \; \;\; \;
m \geq 1
\label{eq3_13}
\end{eqnarray}
the expression for the mean square displacement $\langle x_n^2 \rangle$
can be explicited in the form
\begin{eqnarray}
\langle x_n^2 \rangle & = & \sum_{h=1}^n (1- \pi)
+ 2 \, \sum_{h=1}^n \sum_{k=1}^{h-1} \pi^{h-k} (1- \pi)^2 \nonumber \\
& = & (1-\pi) \, n + 2 \, (1-\pi)^2 \, \sum_{h=1}^n \pi^h \,  \sum_{k=1}^{h-1}
\pi^{-k}
\label{eq3_14}
\end{eqnarray}
Since $\sum_{h=1}^m \alpha^h = (\alpha - \alpha^{m+1})/(1-\alpha)$,
for any real $\alpha$,
it follows that
\begin{eqnarray}
\sum_{h=1}^n \pi^h \,  \sum_{k=1}^{h-1}
\pi^{-k} & =  &\sum_{h=1}^n \pi^h  \, \frac{1/\pi - 1/\pi^h}{1 - 1/\pi}
\nonumber \\
& = & - \frac{\pi}{1-\pi} \left ( \frac{1}{\pi} \sum_{h=1}^n \pi^h - n \right
)
\label{eq3_15}
\end{eqnarray}
For any $\pi \in [0,1)$, $\sum_{h=1}^n \pi^h $ converges to
$1/(1-\pi)$ and consequently,
\begin{eqnarray}
\langle x_n^2 \rangle & = &
(1- \pi ) \, n + \frac{2 \, (1- \pi)^2 \, \pi}{1-\pi} \, n + {\mathcal O}(1)
\nonumber \\
& = & (1-\pi) \, (1+2 \, \pi) \, n + {\mathcal O}(1)
\label{eq3_16}
\end{eqnarray}
Eq. (\ref{eq3_16}) indicates that the effective
self-diffusion coefficient $D_{\rm sd}(\pi)$ for this random walk scheme
in the mean-field approximation equals
\begin{equation}
D_{\rm sd}(\pi) = \frac{(1-\pi) \, (1+2 \, \pi)}{2}
\label{eq3_17}
\end{equation}
The interesting feature of this result is that 
$D_{\rm sd}(\pi)$ displays a non monotonic behavior
as a function of $\pi$: for small $\pi$, $D_{\rm sd}(\pi)$
increases above the value $D_{\rm sd}(0)=1/2$,
while for $\pi \rightarrow 1$, $D_{\rm sd}(\pi) \rightarrow 0$.
This phenomenon is depicted in figures \ref{Fig1} and \ref{Fig2}.
Figure \ref{Fig1} shows the mean square displacement $\sigma_x^2(n)=
\langle x_n^2 \rangle$ vs time $n$ obtained
for  stochastic simulations of eqs. (\ref{eq3_1}) (\ref{eq3_5}),
using  an ensemble of $10^5$ particles,
while figure \ref{Fig2} compares the values
of the self-diffusivity obtained from
the simulations
against the theoretical prediction (\ref{eq3_17}).

Next, consider the same process in the framework of the theory
of GPK processes, still assuming a mean-field approximation.
While there are only two different spin states $\pm 1$ as regards the lattice
model,
 there are four different velocity/spin states in its GPK counterpart:
namely the two states $\pm b_0$ in which particles possess
spin states $\pm 1$, and an effective velocity $\pm b_0$,
and the two ``ghost states'' $O(\pm)$, at which the velocity is vanishing
while the value of the spin state is $\pm 1$.
Let us label these four states with $i=1,..,4$
\begin{table}
\begin{center}
\begin{tabular}{ccc}
\hline
State & Velocity & Spin \\
\hline
1 & $b_0$ & $+$ \\
2 & $-b_0$ & $-$ \\
3 & $0$ & $+$ \\
4 & $0$ & $-$ \\
\hline
\end{tabular}
\end{center}
\caption{Correspondence between the four states of the GPK models and
particle velocity and spin.}
\label{Tab1}
\end{table}
Let $\lambda_0$ be a uniform transition rate.
The stochastic GPK model is thus given by
\begin{equation}
d x (t) = b_{\chi_4(t;\lambda_0 {\bf 1}, {\bf A})} \, dt
\label{eq3_18}
\end{equation}
where the stochastic velocity vector
${\bf b}=(b_i)_{i=1}^4$ corresponds to the
second row of table \ref{Tab1},
\begin{equation}
{\bf b}=
\left (
\begin{array}{c}
b_0 \\
-b_0 \\
0 \\
0
\end{array}
\right )
\label{eq3_19}
\end{equation}
$\chi_4(t;\lambda_0 {\bf 1} , {\bf A})$ is a 4-state finite Poisson process
characterized by a uniform transition rate $\lambda_0$ and by the
transition probability matrix ${\bf A}$ given by
\begin{equation}
{\bf A}=
\left (
\begin{array}{cccc}
 \pi \, (1-\pi) & (1-\pi)^2 & \pi \, (1-\pi) & (1-\pi)^2 \\
(1-\pi)^2 & \pi (1-\pi) & (1-\pi)^2 & \pi (1-\pi) \\
\pi^2 & \pi \, (1-\pi) & \pi^2 & \pi \, (1-\pi)  \\
\pi \, (1-\pi) & \pi^2 &\pi \, (1-\pi) & \pi^2 
\end{array}
\right )
\label{eq3_20}
\end{equation}
Let us clarify the structure of the transition
probability matrix. Consider as initial state,
the state ``$1$''. The transition from this state
to state ``$2$'', corresponding to a moving particle
with opposite velocity, can occur solely if the
initial site does not contain any other particle,
and this happens in the mean-field approximation with
probability $1-\pi$
and the nearest neighbouring site can be reached without
violating the exclusion principle, which occurs
with probability $1-\pi$. The probability $A_{2,1}$ is
therefore equal to $A_{2,1}=(1-\pi)^2$.
The transition from state ``$1$'' to state ``$3$'', corresponding
to a rest particle with the same spin can occur solely
if the initial site contains a particle with oppositive spin,
and the nearest neighboring site is occupied by a particle
with the same spin. Both these events occurs with probability $\pi$,
and are indpendent of each other, so that $A_{3,1}=\pi^2$.
The transition from state ``$1$'' to state ``$4$'' corresponding
to a rest particle with opposite spin, can occurs solely
if the initial state does not contain any other particle (occurring
with probability
$1-\pi$), and the nearest neighbouring site is occupied by
a particle possessing the same spin (occurring
with probability $\pi$), so that $A_{4,1}=\pi \, (1-\pi)$.
As regards $A_{1,1}$, its expression follows from
the probabilistic closure condition $A_{1,1}=1-A_{2,1}-A_{3,1}-A_{4,1}$.
An analogous derivation can be applied to determine all the
other entries of  the matrix ${\bf A}$.

The statistical description of the process   (\ref{eq3_18})
involves four partial probability  density functions
$p_1(x,t)$, $p_2(x,t)$,  $p_3(x,t)$, $p_4(x,t)$
associated with the four states of $\chi_4(t;\lambda_0 {\bf 1}, {\bf A})$,
fulfilling the balance equations
\begin{equation}
\frac{\partial p_i(x,t)}{\partial t} = - b_i \frac{\partial p_i(x,t)}
{\partial x} - \lambda_0 \, p_i(x,t) + \lambda_0 \sum_{j=1}^4 A_{i,j} \,
p_j(x,t)
\label{eq3_21}
\end{equation}
and the overall probability density function is
obviously $p(x,t)=\sum_{i=1}^4 p_i(x,t)$. Let us define
the two probabilistic $2$-vectors
${\bf p}_b(x,t)$, ${\bf p}_0(x,t)$ as
\begin{equation}
{\bf p}_b =
\left (
\begin{array}{c}
p_1 \\
p_2
\end{array}
\right )
\; , \qquad
{\bf p}_0 =
\left (
\begin{array}{c}
p_3 \\
p_4
\end{array}
\right )
\label{eq3_22}
\end{equation}
${\bf p}_b$ is the vector of the partial probability 
density associated with moving states, i.e., with states corresponding
to an effective particle motion, while
${\bf p}_0$ groups together the partial probabilities
pertaining to the rest states.
With this notation, the balance  equations
for the partial probability waves can be compactly expressed
as
\begin{eqnarray}
\frac{\partial {\bf p}_b}{\partial t}
&= & - b_0 {\mathcal L}_x[{\bf p}_b] - \lambda_0 \, {\bf p}_b
+ \lambda_0 \, {\bf A}_1 \, {\bf p}_b + \lambda_0 \, {\bf A}_1 \,
{\bf p}_0 \nonumber \\
\frac{\partial {\bf p}_0}{\partial t}
&= &  -\lambda_0 \, {\bf p}_0 + \lambda_0 \, {\bf A}_2 \, {\bf p}_b + \lambda_0 \, {\bf A}_2 \,
{\bf p}_0 
\label{eq3_23}
\end{eqnarray}
where ${\mathcal L}_x$ is the advection operator
\begin{equation}
{\mathcal L}_x =
\left (
\begin{array}{cc}
\partial/\partial x & 0\\
0 & - \partial/\partial x
\end{array}
\right )
\label{eq3_24}
\end{equation}
and the two $2 \times 2$ matrices ${\bf A}_1$, ${\bf A}_2$ read
\begin{equation}
{\bf A}_1 =
\left (
\begin{array}{cc}
\pi \, (1-\pi) & (1-\pi)^2 \\
(1-\pi)^2 & \pi \, (1-\pi)
\end{array}
\right )
\; ,
\qquad
{\bf A}_2 =
\left (
\begin{array}{cc}
\pi^2 & \pi \, (1-\pi) \\
\pi \, (1- \pi) & \pi^2
\end{array}
\right )
\label{eq3_25}
\end{equation}

The balance equation for the overall probability density
$p(x,t)$ follows from  (\ref{eq3_21}) by summing over the
states (i.e., over the index $i$),
\begin{equation}
\frac{\partial p(x,t)}{\partial t}= - \frac{\partial J_p(x,t)}{\partial x}
\label{eq3_26}
\end{equation}
where $J_p(x,t)=b_0 [p_1(x,t)-p_2(x,t)]$.

Next,  consider the Kac limit of this model, corresponding
to $b_0, \lambda_0 \rightarrow \infty$ keepind fixed
the ratio $b_0^2/2 \lambda_0=D_0$ to a fixed nominal diffusivity $D_0$.
Letting $\lambda_0 \rightarrow \infty$, the second
equation (\ref{eq3_23}) provides the ration
between ${\bf p}_0$ and ${\bf p}_b$,
\begin{equation}
({\bf I} - {\bf A}_2 ) \, {\bf p}_0 = {\bf A}_2 \, {\bf p}_b
\label{eq3_27}
\end{equation}
i.e., ${\bf p}_0= ({\bf I}-{\bf A}_2)^{-1} \, {\bf A}_2 \, {\bf p}_b$.
Indicating with $p_b(x,t)=p_1(x,t)+p_2(x,t)$ and $p_0(x,t)=p_3(x,t)+p_4(x,t)$,
from eq. (\ref{eq3_27}), and from the identity $p(x,t)=p_b(x,t)+p_0(x,t)$
one obtain the relation between $p(x,t)$, $p_b(x,t)$ and $p_0(x,t)$,
namely
\begin{equation}
 p_0(x,t)= \frac{\pi}{1-\pi} \, p_b(x,t) \; , \qquad
p_b(x,t)= (1- \pi) \, p(x,t)
\label{eq3_28}
\end{equation}
that, substituted into the first equation (\ref{eq3_23}),
yields a balance equation involving solely the partial probability
density associated with moving particles
\begin{eqnarray} 
\frac{\partial {\bf p}_b}{\partial t} &= &
- b_0 \, {\mathcal L}_x[{\bf p}_b] - \lambda_0 [ {\bf I} -{\bf A}_1
- {\bf A}_1 \, ({\bf I}- {\bf A}_2 )^{-1} \, {\bf A}_2 ] \, {\bf p}_b
\nonumber \\
& = & - b_0 \, {\mathcal L}_x[{\bf p}_b] - \lambda_0 {\boldsymbol \Lambda}_{\rm eff}(\pi) \,
{\bf p}_b
\label{eq3_29}
\end{eqnarray}
After some elementary algebra, the matrix ${\boldsymbol \Lambda}_{\rm eff}(\pi)$
takes the form
\begin{equation}
{\boldsymbol \Lambda}_{\rm eff}(\pi) =
\left (
\begin{array}{cc}
\ell(\pi) & -\ell(\pi) \\
-\ell(\pi) & \ell(\pi)
\end{array}
\right ) \; , \qquad
\ell(\pi) = \frac{1}{1+ 2 \, \pi}
\label{eq3_30}
\end{equation}
With respect to the partial probability densities associated with
moving states $(p_1,p_2)$, the statistical description of
the process reduces to a classical Poisson-Kac model
\begin{eqnarray}
\frac{\partial p_1}{\partial t} & = & -b_0 \frac{\partial p_1}{\partial x}
- \lambda_0 \, \ell(\pi) \, [p_1-p_2] \nonumber \\
\frac{\partial p_2}{\partial t} & = & b_0 \frac{\partial p_2}{\partial x}
+\lambda_0 \, \ell(\pi) \, [p_1-p_2] \nonumber 
\label{eq3_31}
\end{eqnarray}
the Kac limit of which provides the expression for
probability flux $J_d(x,t)$ entering the balance equation (\ref{eq3_26})
\begin{eqnarray}
J_d(x,t) &=& - \frac{b_0}{2 \, \lambda_0} \, \frac{1}{\ell(\pi)}
\, \frac{\partial p_b(x,t)}{\partial x}
 =  - D_0 \, \frac{1-\pi}{\ell(\pi)} \, \frac{\partial p(x,t)}{\partial x}
\nonumber \\
& = & - D_0 (1 -\pi) \, (1+2 \, \pi) \, \frac{\partial p(x,t)}{\partial x}
\label{eq3_32}
\end{eqnarray}
Setting the nominal diffusivity $D_0=1/2$, one obtains
from eq. (\ref{eq3_32}) the expression for the mean-field
self-diffusivity (\ref{eq3_17}) derived from the original
stochastic model.
Several observations deserve some attention:
\begin{itemize}
\item in the derivation of eq. (\ref{eq3_32}) we have first
considered the limit for $\lambda_0 \rightarrow \infty$ 
in the second equation  (\ref{eq3_23}) for the
probability densities ${\bf p}_0(x,t)$ for the non-moving
particles, and the result obtained is then substituted back in
the first equation (\ref{eq3_23}) for ${\bf p}_b(x,t)$,
deriving the self-diffusion from the Kac limit of this
equation. We have use this, more physically oriented,
approach to obtain $D_{\rm sd}(\pi)$ in order
to derive eq. (\ref{eq3_31}) corresponding
to the quasi steady-state approximation for the
dynamics of the partial probability density functions
associated with non-moving particles.
If one perform simultaneously the Kac limit, (i.e.,
$\lambda_0, b_0 \rightarrow \infty$, keeping fixed the
nominal diffusiviy $D_0=b_0^2/2 \lambda_0$) one still
obtains eq. (\ref{eq3_32}).
\item The analysis of the above problem involving interacting
particles throu\-gh an exclusion principle indicates
that, once the microdynamics of the interacting particles
has been specified (in the present case within the
mean-field approximation), it is rather straightforward to
define and derive the corresponding stochastic GPK model,
in the present case eq. (\ref{eq3_18}), specified
by the number $N$ of GPK states, by the stochastic velocities
$b_i$  of each state $i=1,\dots,N$, by the
transition rate vector ${\boldsymbol \Lambda}$ and by
the transition probability matrix ${\bf A}$.
\item Observe that the number of states in the GPK model may be different,
and in general greater than the number of spin states of the
original system. In the present case, the
number of different spin configurations is $2$, while $N=4$.
This is because two additional states are required to discriminate
between moving and non-moving particles in order
to account for the exclusion principle;
\item As discussed with the aid of the present case study,
it is fairly easy to derive the structure of ${\boldsymbol \Lambda}$
and ${\bf A}$ and their dependence on the partial probability
density functions (in the presence case on the 
concentration $\pi$, since the simpler case of a
mean-field approximation is considered) from the
rules of particle interaction. The analysis
developed in this Section is limited to the mean-field case. The
general problem is treated in the next Section.
\item Given the stochastic GPK model (in the present case eq.
(\ref{eq3_18})), the hydrodynamic limit of this model
follows directly from GPK theory, in the present case eq.
(\ref{eq3_21}). Out of it, the Kac-limit of the latter,
provides the classical parabolic transport model.
Therefore, and this represents a very powerful by-product of
GPK theory, there are several classes of hydrodynamic
limits of the same interacting particle systems, depending,
once $D_0$ is fixed, on the characteristic time scales
of the stochastic process, i.e., essentially on the value
of $\lambda$. In some cases, due to the presence of
particle interactions, while the hyperbolic
hydrodynamic limit exists, the Kac limit of the
corresponding model could  not exist.
\end{itemize}

\subsection{Simple exclusion random walk}
\label{sec3_2}

Let us consider another classical exclusion random walk without
spin. In this
model, particles on a lattice move towards the nearest
neighbouring site (with equal probability towards the
left or right neighboring site) solely
if no other  particle is simultaneosly occupying it.

In the mean-field approximation, indicating with $\pi$
particle concetration, the random walk model takes
the form
\begin{equation}
x_{n+1}= x_n + r_{n+1} \, \eta_{n+1}
\label{eq3_33}
\end{equation}
where the random variables $r_{n+1}$, $\eta_{n+1}$ are specified
by
\begin{equation}
r_{n+1}=
\left \{ 
\begin{array}{lll}
-1 & & \mbox{Prob} \; \, 1/2 \\
1 & & \mbox{Prob} \; \, 1/2 
\end{array}
\right .
\; ,
\qquad
\eta_{n+1} =
\left \{
\begin{array}{lll}
0 & & \mbox{Prob} \; \, \pi \\
1 & & \mbox{Prob} \; \, (1-\pi)
\end{array}
\right .
\label{eq3_34}
\end{equation}
where $r_h$ and $\eta_k$ are uncorrelated with
each other, $\langle r_h \rangle =0$,
$\langle r_h \, r_k \rangle =\delta_{h,k}$ and
$\eta_h$ satisfy eq. (\ref{eq3_7}).
Consequently, starting from $x_0=0$,
\begin{equation}
x_n= \sum_{h=1}^n r_h \, \eta_h 
\label{eq3_35}
\end{equation},
and $\langle x_n\rangle=0$ while
for the mean square displacement
\begin{equation}
\langle x_n^2 \rangle = \sum_{h=1}^n \sum_{k=1}^n \langle r_h \,
r_k \rangle \, \langle \eta_h \, \eta_k \rangle
= \sum_{h=1}^n \langle \eta_h^2 \rangle = (1-\pi) \, n
\label{eq3_36}
\end{equation}
Thus, for the self diffusion coefficient of a tagged particle
one obtains, in the mean-field approximation,
\begin{equation}
D_{\rm sd}(\pi)= \frac{1-\pi}{2} 
\label{eq3_37}
\end{equation}
Next, consider the GPK modeling. The GPK
version of the process involves three states: state ``1''
corresponding to particles moving forward along the $x$-axis,
state ``2'' to particles moving backward, and state ``3'' corresponding
to resting particles, that do not perform any  motion due to
the exclusion principle. Indicating with $b_0$ and
$\lambda_0$ the characteristic velocity and transition
rate of the GPK process, it follows that
\begin{equation}
{\bf b}=
\left (
\begin{array}{c}
b_0 \\
-b_0 \\
0
\end{array}
\right ) \; ,
\qquad
{\boldsymbol \Lambda}
=  \lambda_0 \, \left (
\begin{array}{c}
1 \\
1 \\
1
\end{array}
\right )
\label{eq3_38}
\end{equation}
and the stochastic GPK version of the model
is formally analogous to eq. (\ref{eq3_18}), namely
\begin{equation}
d x (t) = b_{\chi_3(t;\lambda_0 {\bf 1}, {\bf A})} \, dt 
\label{eq3_39}
\end{equation}
where the transition probability matrix ${\bf A}$
depends on the mean-field concentration $\pi$ and is  given by
\begin{equation}
{\bf A}
= \left (
\begin{array}{ccc}
\frac{1-\pi}{2} & \frac{1-\pi}{2} & \frac{1-\pi}{2} \\
\frac{1-\pi}{2} & \frac{1-\pi}{2} & \frac{1-\pi}{2} \\
\pi & \pi & \pi
\end{array}
\right )
\label{eq3_40}
\end{equation}
The statistical description of eq. (\ref{eq3_39}) involves
three partial probability densities $p_1(x,t)$, $p_2(x,t)$
and $p_3(x,t)$, where the latter corresponds to the density of
resting particles. The overall
probability density is $p(x,t)=\sum_{i=1}^3 p_i(x,t)$,
and indicate with $p_b(x,t)=p_1(x,t)+p_2(x,t)$
the probability density function of the moving particles.
The balance equations for  the partial densities,
accunting for eqs. (\ref{eq3_38}) and (\ref{eq3_40})
are
\begin{eqnarray}
\frac{\partial p_1(x,t)}{\partial t} & = & - b_0 \frac{\partial p_1(x,t)}{\partial x} - \lambda_0 \, p_1(x,t) + \lambda_0 a(\pi) \, p(x,t)
\nonumber \\
\frac{\partial p_2(x,t)}{\partial t} & = &  b_0 \frac{\partial p_2(x,t)}{\partial x} - \lambda_0 \, p_2(x,t) + \lambda_0 a(\pi) \, p(x,t)
\label{eq3_41} \\
\frac{\partial p_3(x,t)}{\partial t} & = &
- \lambda_0 \, p_3(x,t) + \lambda_0 \, \pi \, p(x,t)
\nonumber 
\end{eqnarray}
where $a(\pi)=(1-\pi)/2$,
from which one obtains  that the conservation equation
for the overall density is still expressed by eq. (\ref{eq3_26})
with $J_d(x,t)= b_0 \, [p_1(x,t)-p_2(x,t)]$.
As regards the probability flux $J_d(x,t)$, from the
first two equations  (\ref{eq3_41}) one obtains
\begin{equation}
\frac{\partial J_d(x,t)}{\partial t} = - b_0^2 \,
\frac{\partial p_b(x,t)}{\partial x} - \lambda_0 \,  J_d(x,t)
\label{eq3_42}
\end{equation}
In the Kac limit, eq. (\ref{eq3_42}) provides
\begin{equation}
J_d(x,t)= - 2 \, D_0 \, \frac{\partial p_b(x,t)}{\partial x}
\label{eq3_43}
\end{equation}
where, as usual, $D_0=b_0^2/2 \lambda_0$ corresponds to
the nominal diffusivity of the GPK scheme.
In the Kac limit, from the third equation (\ref{eq3_41})
one obtains $p_3(x,t) = \pi \, p(x,t)$, thus
\begin{equation}
p_b(x,t) = (1- \pi) \, p(x,t)
\label{eq3_44}
\end{equation}
which inserted into eq. (\ref{eq3_43}) provides
\begin{equation}
J_d(x,t)= - 2 \, D_0 \,  (1- \pi) \, \frac{\partial p(x,t)}{\partial x}
\label{eq3_45}
\end{equation}
which implies for the self-diffusion $D_{\rm sd}(\pi)=2 \, D_0
(1-\pi)$ that coincides with eq. (\ref{eq3_37}), by
setting the nominal diffusivity equal to $D_0=1/4$.

\subsection{TASEP model}
\label{sec3_3}
To conclude, let us consider another simple and paradigmatic
example, namely the Totally Asymmetric Simple Exclusion Process (TASEP)
on the real line. In this model, particles move solely in the
forward direction satisfying an exclusion principle, corresponding to
one particle per site, at most.
The mean field dynamics of TASEP, letting $\pi$ be the mean-field
particle concentration, is described by the dynamics
\begin{equation}
x_{n+1}=x_n+ \xi_{n+1}
\label{eq3_46}
\end{equation}
where the random variables $\xi_{h}$ are uncorrelated
with each other and described statistically by
\begin{equation}
\xi_{n+1}=
\left \{
\begin{array}{cl}
0 & \mbox{Prob} \; \, \pi \\
1 & \mbox{Prob} \; \, 1-\pi
\end{array}
\right .
\label{eq3_47}
\end{equation}
Consequently,
\begin{equation}
\langle \xi_h \, \xi_k \rangle
= \left \{
\begin{array}{ccl}
1-\pi & & h=k \\
(1-\pi)^2 & &  h \neq k
\end{array}
\right .
\label{eq3_48}
\end{equation}
If $x_0=0$, the integral representation of the
dynamics is $x_n=\sum_{h=1} \xi_h$, thus
\begin{equation}
\langle x_n \rangle = \sum_{h=1}^n \langle \xi_h \rangle = (1- \pi) \, n
\label{eq3_49}
\end{equation}
and
\begin{equation}
\langle x_n^2 \rangle = \sum_{h=1}^n \sum_{k=1}^n \langle \xi_h \, \xi_k \rangle
= (1-\pi)^2 \, n^2 + \pi \, (1-\pi) \, n
\label{eq3_50}
\end{equation}
The mean square displacement attains the expression $\sigma_x^2(n)
= \langle x_n^2 \rangle - \langle x_n \rangle^2= \pi \, (1-\pi) \, n$,
thus the mean-field self-diffusion coefficient is given by
\begin{equation}
D_{\rm sd}(\pi) = \frac{\pi \, (1- \pi)}{2}
\label{eq3_51}
\end{equation}
It vanished both for $\pi=0$ (infinite dilution) and for $\pi=1$
corresponding to total exclusion. In both cases the dynamics is strictly
(and trivially) deterministic.

Let us analyze the GPK formulation of TASEP. It implies
the occurrence of two state: state ``1'' which is the
mobile state, and state ``2'' which is the stationary (non-moving)
state. Correspondingly, $b_1=b_0$, and $b_2=0$.
The transition rates are uniform and equal to $\lambda_0$.
As regards the transition probability matrix, TASEP dynamics
indicates the following dependence on the mean field concentration $\pi$
\begin{equation}
{\bf A} =
\left (
\begin{array}{cc}
1- \pi & 1- \pi \\
\pi & \pi
\end{array}
\right )
\label{eq3_52}
\end{equation}
The process is described stastically by the two
partial probability density functions satisfying the
hyperbolic equations
\begin{eqnarray}
\frac{\partial p_1(x,t)}{\partial t} & = & - b_0 \, \frac{\partial p_1(x,t)}{\partial x}
- \lambda_0 \, \pi \, p_1(x,t) + \lambda_0 \, (1-\pi) \, p_2(x,t)
\nonumber \\
\frac{\partial p_0(x,t)}{\partial t} & = & \lambda_0 \, \pi \, p_1(x,t) - \lambda_0 \, (1-\pi)
\, p_2(x,t)
\label{eq3_53}
\end{eqnarray}
In the limit $\lambda_0 \rightarrow \infty$,
 $\lambda_0^{-1} \partial p_2/\partial t=0$,
thus the second equation (\ref{eq3_53}) provides
\begin{equation}
p_2(x,t)= \frac{\pi}{1-\pi} \, p_1(x,t) \; ,
\qquad
p_1(x,t) = (1-\pi) \, p(x,t)
\label{eq3_54}
\end{equation}
where $p(x,t)=p_1(x,t)+p_2(x,t)$ is the overall concentration,
the dynamics of which is given by
\begin{equation}
\frac{\partial p(x,t)}{\partial t}= - b_0 \, \frac{\partial p_1(x,t)}{\partial x}
\label{eq3_55}
\end{equation}
Inserting in it the expression (\ref{eq3_54}),
 obtained for $\lambda_0 \rightarrow \infty$,
 one finally arrive to the hyperbolic
model for $p(x,t)$
\begin{equation}
\frac{\partial p(x,t)}{\partial t}= - b_0 \, (1-\pi) \,
\frac{\partial p(x,t)}{\partial x}
\label{eq3_56}
\end{equation}
providing an effective mean velocity $v_{\rm eff}$ equal to
\begin{equation}
v_{\rm eff}= b_0 \, (1- \pi)
\label{eq3_57}
\end{equation}
Observe that eqs. (\ref{eq3_56})-(\ref{eq3_57}) does not
correspond to any Kac limit, but solely to the limit
of infinitely fast recombination kinetics ($\lambda_0 \rightarrow
\infty$).

In order to extract from the GPK process defined
statistically by eq. (\ref{eq3_53}) the value for the effective
diffusivity and to perform a  Kac limit of the process,
let us consider TASEP dynamics in the inertial frame moving with
the effective velocity $v_{\rm eff}$.
Let $x^\prime$ be the position coordinate in this moving reference system
\begin{equation}
x^\prime = x - b_0 \, (1- \pi) \, t
\label{eq3_58}
\end{equation}
In the moving system, let $p_1^\prime(x^\prime,t)$, $p_2^\prime(x^\prime,t)$
the two partial probability densities characterized by the
velocities
\begin{equation}
b_1^\prime = b_0- b_0 \, (1- \pi) = b_0 \, \pi \; ,
\qquad
b_2^\prime = - b_0 \, (1- \pi)
\label{eq3_59}
\end{equation}
which satisfty the balance equations
\begin{eqnarray}
\frac{\partial p_1^\prime(x^\prime,t)}{\partial t} & = &
= - b_0 \, \pi \, \frac{\partial p_1^\prime(x^\prime,t)}{\partial x^\prime} -
\lambda_0 \, \left [ \pi \, p_1^\prime(x^\prime,t) - (1- \pi) \, p_2^\prime(x^\prime,t) \right ] \nonumber \\
\frac{\partial p_2^\prime(x^\prime,t)}{\partial t} & = &
=  b_0 \, ( 1- \pi) \, \frac{\partial p_2^\prime(x^\prime,t)}{\partial x^\prime} +
\lambda_0 \, \left [ \pi \, p_1^\prime(x^\prime,t) - (1- \pi) \, p_2^\prime(x^\prime,t) \right ] \nonumber \\
\label{eq3_60}
\end{eqnarray}
The stochastic GPK process
\begin{equation}
d x^\prime(t) = b_{\chi_2(t;\lambda {\bf 1}, {\bf A}(\pi)}(\pi) \, dt
\label{eq3_61}
\end{equation}
 associated with the statistical description (\ref{eq3_60}),
where $b_i$, $i=1,2$, are given by eq. (\ref{eq3_58}), and the
transition probability matrix ${\bf A}(\pi)$  by eq. (\ref{eq3_52})
will be referred to as the zero-bias TASEP model.
The balance equation for the overall probability
density $p^\prime(x^\prime,t)=p_1^\prime(x^\prime,t)+ p_2^\prime(x^\prime,t)$
is obviously given by
\begin{equation}
\frac{\partial p^\prime(x^\prime,t)}{\partial t} = - \frac{\partial J_p^\prime(x^\prime,t)}{\partial x^\prime} 
\label{eq3_62}
\end{equation}
where
$J_d= b_0 (\pi \, p_1^\prime- (1-\pi) \, p_2^\prime)$,
the evolution of which is given by
\begin{equation}
\frac{\partial J_p^\prime(x^\prime,t)}{\partial t} =
- b_0^2 \, \frac{\partial }{\partial x^\prime}
\left [ \pi^2 \, p_1^\prime(x^\prime,t) + (1- \pi)^2 \,
p_2^\prime(x^\prime,t)  \right ] - \lambda_0 \, J_dì\prime(x^\prime,t)
\label{eq3_63}
\end{equation}
From the definition of $p^\prime$ and $J_p^\prime$ in terms
of $p_1^\prime$, $p_2^\prime$ the inverse relations follow
\begin{equation}
p_1^\prime = (1-\pi) \, p^\prime + \frac{J_p}{b_0}
\; , \qquad
p_2^\prime = \pi \, p^\prime - \frac{J_p}{b_0}
\label{eq3_64}
\end{equation}
Therefore, in the Kac limit, $b_0, \, \lambda_0 \rightarrow \infty$
it follows that
\begin{eqnarray}
J_p^\prime  &= &- 2 \, D_0 \ \left ., \frac{\partial }{\partial x^\prime}
\left [ \pi^2 \, p_1^\prime + (1-\pi)^2 \, p_2^\prime \right ]
\right |_{p_1^\prime= (1-\pi) \, p^\prime \, , \; \, 
p_2^\prime= \pi \, p^\prime} \nonumber \\
& = & - 2\,  D_0 \, \pi \, (1- \pi) \, \frac{\partial p^\prime}{\partial x^\prime}
\label{eq3_65}
\end{eqnarray}
and therefore the effective self-diffusion coefficient
is given by $D_{\rm sd}(\pi)=  2 \, D_0 \, \pi \, (1- \pi)$,
consistently with the expression (\ref{eq3_51}) deriving
from the lattice representation of TASEP, by choosing
$D_0=1/4$ for the nominal diffusivity.

\section{Dynamic nonlinear models}
\label{sec4}

In the previous Section we have  considered exclusively the
mean-field approximation corresponding to the motion of tagged
particles in the mean-field characterized by a fixed concentration.
The mean-field approximation has been introduced essentially
in order: (i)  to connect a generic physical model of interacting particles,
with its corresponding GPK process, (ii) to show how the latter can be
easily built up from lattice dynamics, and (iii) to show
the existence of several hydrodynamic limits.

In considering the dynamics of an interacting particle system,
it is rather clear that the mean field approximation
is insufficient to provide a correct description of its evolution
for the simple reason that the average concentration
$\pi$ (introduced in the mean-field modeling) cannot
be regarded as constant as it is a function of both
time and space coordinates.

From the analysis developed in the previous section,
the mean-field GPK model of a system
of interacting particles is defined by
\begin{equation}
d x(t) = b_{\chi_N(t;{\boldsymbol \Lambda}(\pi),{\bf A}(\pi)}(\pi) \, dt
\label{eq4_1}
\end{equation}
where the parameters defining the process, namely  the
stochastic velocities $b_i(\pi)$ characteristic of the $i$-th
state, $i=1,\dots,N$, the transition rate vector ${\boldsymbol \Lambda}(\pi)$
and the transition probability matrix ${\bf A}(\pi)$, depend in
general on the mean-field concentration $\pi$. 
Observe that eq. (\ref{eq4_1}) implicitly assume that
no external biasing fields are present, as for 
the cases treated in Section \ref{sec3}
(the TASEP model considered in paragraph \ref{sec3_3} is
obviously characterized by an internal drift, but the GPK model
of the process refers to its description in a reference system moving
at the effective velocity of the process). If an external velocity
field $v(x)$ is present, it can be included into eq. (\ref{eq4_1})
by adding to as a  drift $v(x(t)) \, dt$ in the equation for $d x(t)$.

In order to consider the proper dynamics of a system of particles,
the mean-field formulation (\ref{eq4_1})
should be replaced by a nonlinear stochastic dynamics of the
form
\begin{equation}
d x(t) = \widehat{b}_{\chi_N(t;\widehat{\boldsymbol \Lambda}({\bf p}(x(t),t)),\widehat{\bf A}({\bf p}(x,t))}({\bf p}(x,t)) \, dt
\label{eq4_2}
\end{equation}
where $\widehat{b}_i$, $\widehat{\boldsymbol \Lambda}=(\widehat{\lambda}_1,\dots,\widehat{\lambda}_N)$ and
$\widehat{\bf A}$ depend  on the entire system of 
partial probability densities $p_i(x,t)$, $i=1,\dots,N$  
characterizing the process,
i.e., on the vector-valued probability density
${\bf p}(x,t)=(p_1(x,t),\dots,p_N(x,t))$.
The explicit expression for $\widehat{b}_i$, $\widehat{\boldsymbol \Lambda}$
and $\widehat{\bf A}$, can be  derived from the mechanics
of particle interaction, similarly to what developed
in Section \ref{sec3} for the mean-field case.
Moreover, a self-consistency condition should be fulfilled,
namely that if all the $p_i$ equal $\pi/N$ that these
quantities should coincide with the corresponding mean-field
counterparts, i.e.,
\begin{equation}
\left . \widehat{b}_i({\bf p})= \right |_{p_h=\pi/N, \; h=1,\dots n}=
b_i(\pi) \; , \qquad i=1,\dots,N
\label{eq4_3}
\end{equation}
and analogous for the remaining quantities.
Since a slightly different normalization has been
adopted in the case of the fermionic model addressed
in paragraph \ref{sec3_1}, where $\pi$ is the
mean-field concentration associated to a given spin-value
(either $+1$ or $-1$) the consistency condition
(\ref{eq4_3}) still applies to this case
with $p_h= 2 \,  \pi/N$, since the
spin states are two.

Henceforth, for simplifying the notation, we will
indicate the ``hatted'' quantity, say $\widehat{b}_i$,
solely with the bare letters and superscript, e.g. $b_i$.

Equation (\ref{eq4_2}) is a nonlinear stochastic model,
in which particle stochastic motion depends on the
collective state of the system at any time $t$.
It should be interpreted a la McKean \cite{mckean,nfp}
and leads to nonlinear balance equations for the
partial probability densities \cite{gpk3} which
can be explicited as
\begin{eqnarray}
\frac{\partial p_i(x,t)}{\partial t}
&= & - \frac{\partial }{\partial x} \left ( b_i({\bf p}(x,t)) \, p_i(x,t)
\right ) - \lambda_i({\bf p}(x,t)) 
\nonumber \\
& + & \sum_{j=1}^N A_{i,j}({\bf p}({\bf x},t)) \, \lambda_j({\bf p}(x,t))
\, p_j(x,t)
\label{eq4_4}
\end{eqnarray}
Below, we analyze the systems addressed in Section \ref{sec3}
in order to extract their proper dynamic characterization
and to derive the Kac limit of eq. (\ref{eq4_4}),
proceeding in the reverse order, namely from the simpler (TASEP)
to the most elaborate fermionic model addressed in paragraph \ref{sec3_1}

\subsection{TASEP in the moving reference frame}
\label{sec4_1}

Consider again the TASEP model in the moving reference
frame addressed in paragraph \ref{sec3_3} eqs. (\ref{eq3_59})-(\ref{eq3_61}),
dropping the  prime superscript ($\prime$) for notational simplicity.
In the TASEP model, the mean-field concentration $\pi$ corresponds,
in a fully dynamic description of the process to $p(x,t)=p_1(x,t)
+p_2(x,t)$. Consequently the stochastic velocity vector
of the two-state GPK process are given by
\begin{equation}
b_1= b_0 \, p(x,t) \; , \qquad
b_2= -b_0 [1-p(x,t)]
\label{eq4_5}
\end{equation}
the transition rates are uniform, i.e, ${\boldsymbol \Lambda}=(\lambda_0,
\lambda_0)$, and the transition probability matrix ${\bf A}({\bf p})$
takes the form
\begin{equation}
{\bf A}({\bf p})
= \left (
\begin{array}{cc}
1-p & 1-p \\
p & p
\end{array}
\right )
\label{eq4_6}
\end{equation}
Therefore, the balance equations for the partial probability densities
(partial concentrations) are given by
\begin{eqnarray}
\frac{\partial p_1}{\partial t} & = & - b_0 \frac{\partial ( p \, p_1)}{\partial
x} - \lambda_0 \left [ p \, p_1 -(1-p) \, p_2 \right ] \nonumber \\
\frac{\partial p_2}{\partial t} & = &  b_0 \frac{\partial [(1- p) \, p_1]}{\partial
x} + \lambda_0 \left [ p \, p_1 -(1-p) \, p_2 \right ]
\label{eq4_7}
\end{eqnarray}
Eq.  (\ref{eq4_7}) already represents a hydrodynamic limit of
the zero-bias TASEP model, characterized by a finite value of
the characteristic transition rate $\lambda_0$ and by the
diffusivity $D_0$, as $b_0$ is related to $\lambda_0$ and $D_0$ by
the relation $b_0=\sqrt{2 \, D_0 \, \lambda_0}$.
Summing together the two equations in (\ref{eq4_7}) the
dynamics of the overall probability density follows
\begin{equation}
\frac{\partial p}{\partial t} = - \frac{\partial }{\partial x}
\left [ b_0 \, (p \, p_1 - (1-p) \, p_2) \right ] =
- \frac{\partial }{\partial x} \left [ b_0 \, ( p^2 - p_2) \right ]
\label{eq4_8}
\end{equation}
Therefore, the probability flux $J_p$ is given by $J_p=b_0 \, (p^2-p_2)$.
Taking its time derivative
\begin{eqnarray}
\frac{\partial J_p}{\partial t} & = & b_0  \, \left ( 2 \, p \, \frac{\partial p}{\partial t} - \frac{\partial p_2}{\partial t} \right ) \nonumber \\
& = & - 2 \, b_0 \, p \,  \frac{\partial J_p} {\partial x} -
b_0^2 \,  \frac{\partial [(1-p) \, p_2]}{\partial x} - \lambda_0 \, b_0 \, J_p
\label{eq4_9}
\end{eqnarray}
which, in the Kac limit, takes the form
\begin{equation}
J_p = - 2 \, D_0 \,   \frac{\partial [(1-p) \, p_2]}{\partial x} 
\label{eq4_10}
\end{equation}
The first eq.  (\ref{eq4_7}) can be rewritten as
\begin{equation}
\frac{1}{\lambda_0} \, \frac{\partial p_1}{\partial t}
= - \frac{2 \, D_0}{b_0} \, \frac{\partial (p \, p_1)}{\partial x}
- \left  [p \, p_1 -(1-p) \, p_2  \right ]
\label{eq4_11}
\end{equation}
which implies in the Kac limit
\begin{equation}
p_1 = \frac{1-p}{p} \, p_2
\label{eq4_12}
\end{equation}
thus $p=(1+(1-p)/p) \, p_2$, leading to
\begin{equation}
p_2=p^2
\label{eq4_13}
\end{equation}
Substituting this result into eq. (\ref{eq4_10}),
the probability flux becomes $J_p= -2 \, D_0 \, \partial [p^2 \, (1-p)]/\partial x$, and correspondingly the Kac limit of the dynamic TASEP model (in
the zero-bias case) provides the nonlinear diffusion equation
\begin{equation}
\frac{\partial p}{\partial t} = 2 \, D_0 \, \frac{\partial^2 \left [ 
p^2 \, (1-p) \right ]}{\partial x^2}
\label{eq4_14}
\end{equation}

\subsection{Simple exclusion random walk}
\label{sec4_2}
The dynamic analysis of the simple exclusion random walk,
that
in mean-field approximation  has been analyzed in paragraph
\ref{sec3_2}, is conceptually identical to the previous case,
and is completely resolved by identifying the mean-field effective
concentration $\pi$ entering the transition probability matrix ${\bf A}$
eq. (\ref{eq3_52}) with the overall probability density $p=p_1+p_2+p_3$.
Thus, the statistical characterization of the process is
characterized by hyperbolic system
\begin{eqnarray}
\frac{\partial p_1}{\partial t} & = & - b_0 \, \frac{\partial p_1}{\partial
x} - \lambda_0 \, p_1 + \lambda_0 \, a(p) \, p
\nonumber \\
\frac{\partial p_2}{\partial t} & = &  b_0 \, \frac{\partial p_2}{\partial
x} - \lambda_0 \, p_2 + \lambda_0 \, a(p) \, p
\label{eq4_15} \\
\frac{\partial p_3}{\partial t} & = & -\lambda_0 \, p_3  + \lambda_0 \, p^2
\nonumber
\end{eqnarray}
The overall balance for $p(x,t)$ is still expressed by eq. (\ref{eq3_26}),
 the probability flux $J_p(x,t)$ is a solution of eq. (\ref{eq3_42}) where
$p_b=p_1+p_2$.
From the third equation (\ref{eq4_15}), in the limit for $\lambda_0 \rightarrow
\infty$, $p_3=p^2$, thus
\begin{equation}
p=p_b+p_3  \Rightarrow p_b=p \, (1-p)
\label{eq4_16}
\end{equation}
Consequently, from eqs. (\ref{eq3_42}), (\ref{eq4_16}) follows
that $J_p= - 2 \, D_0 \partial [p(1-p)]/\partial x$,
and the Kac limit for the simple exclusion process
attains the form
\begin{equation}
\frac{\partial p}{\partial t} = 2 \, D_0 \, \frac{\partial^2 \left [ 
p \, (1-p) \right ]}{\partial x^2}
\label{eq4_17}
\end{equation}

\subsection{Fermionic random walk with exclusion}
\label{sec4_3}

The dynamic characterization of the fermionic process described
in paragraph \ref{sec3_1} is slightly more difficult than 
the cases  so far considered due to the existence of two
spin states. The concentrations of particles possessing spin $+1$
and $-1$ are given by
\begin{equation}
p_+=p_1+p_3 \; , \qquad p_-=p_2+p_4
\label{eq4_18}
\end{equation}
respectively, and both these quantities equal $\pi$ in the
mean-field approximation.

By considering carefully the exclusion rules characterizing
this process, the dynamic representation of the
transition probability matrix of the associated  GPK model
is given by
\begin{equation}
{\bf A}({\bf p}) =
\left (
\begin{array}{cccc}
p_-  (1-p_+) & (1-p_+)^2 & p_- (1-p_+) & (1-p_+)^2  \\
(1-p_-)^2 & p_+  (1-p_-) & (1-p_-)^2  & p_+ (1-p_-) \\
p_+ \, p_- &  p_+  (1-p_+) & p_+ \, p_- &  p_+ (1-p_+)  \\
p_-  (1-p_-) &   p_+ \, p_- & p_-  (1-p_-)   & p_+ \, p_- 
\end{array}
\right )
\label{eq4_19}
\end{equation}
while ${\bf b}=(b_0,-b_0,0,0)$ and ${\boldsymbol \Lambda}=\lambda_0 (1,1,1,1)$.
From the expression of the quantities describing the GPK process,
the balance equations for the partial densities can be straighforwardly
derived. The Kac limit of this model can be obtained following
the same approach applied in paragraph \ref{sec3_1} for the
mean-field analys, and is not repeated here.
One obtains for $p_b=p_1+p_2$
\begin{equation}
p_b= \left ( 1 - \frac{p}{2} \right ) \, p
\label{eq4_20}
\end{equation}
and for the probability flux
\begin{equation}
J_d = - D_0 \,(1+p) \, \frac{\partial p_b}{\partial x}
= - D_0 \, (1+p) \, \frac{\partial }{\partial x}
\left [ \left ( 1 - \frac{p}{2} \right ) \, p \right ]
\label{eq4_21}
\end{equation}
Consequently the balance equation for the $p(x,t)$ in the Kac limit
reads
\begin{equation}
\frac{\partial p}{\partial t}= D_0 \, \frac{\partial }{\partial x}
\left [ (1+p) \, \frac{\partial }{\partial x} \left ( p- \frac{p^2}{2}
\right ) \right ]
\label{eq4_22}
\end{equation}
Observe that this result is consistent with the mean-field analysis,
as the mean-field concentration  $\pi$ corresponds to $p/2$.

\subsection{General observations}
\label{sec4_4}

From the analysis developed above it follows that the Kac limit
of the exclusion processes analyzed leads to  nonlinear
diffusion equations of the form
\begin{equation}
\frac{\partial p(x,t)}{\partial t}= \frac{\partial^2 W(p(x,t))}{\partial x^2}
\label{eq4_23}
\end{equation}
where  $W(p)$ is a function of  the overall concentration $p(x,t)$
and depends on
the specific model considered, as reviewed in table \ref{Tab2}.

\begin{figure}[!]
\begin{center}
\epsfxsize=8.cm
\epsffile{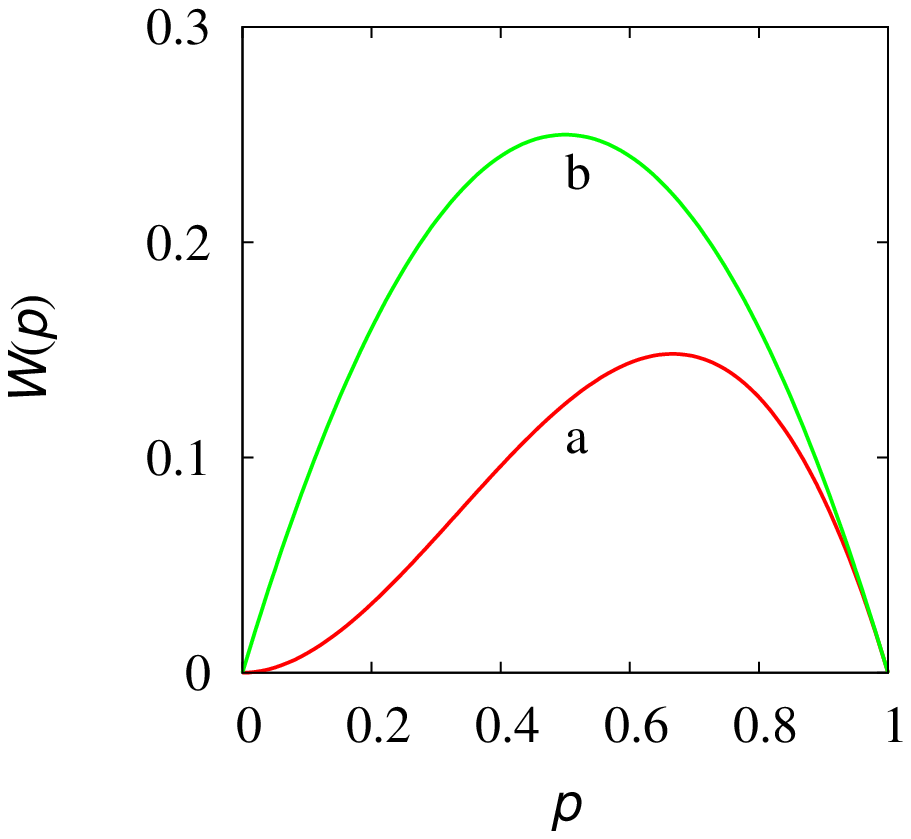} \hspace{-1.5cm} {\large (a)} \\
\epsfxsize=8.cm
\epsffile{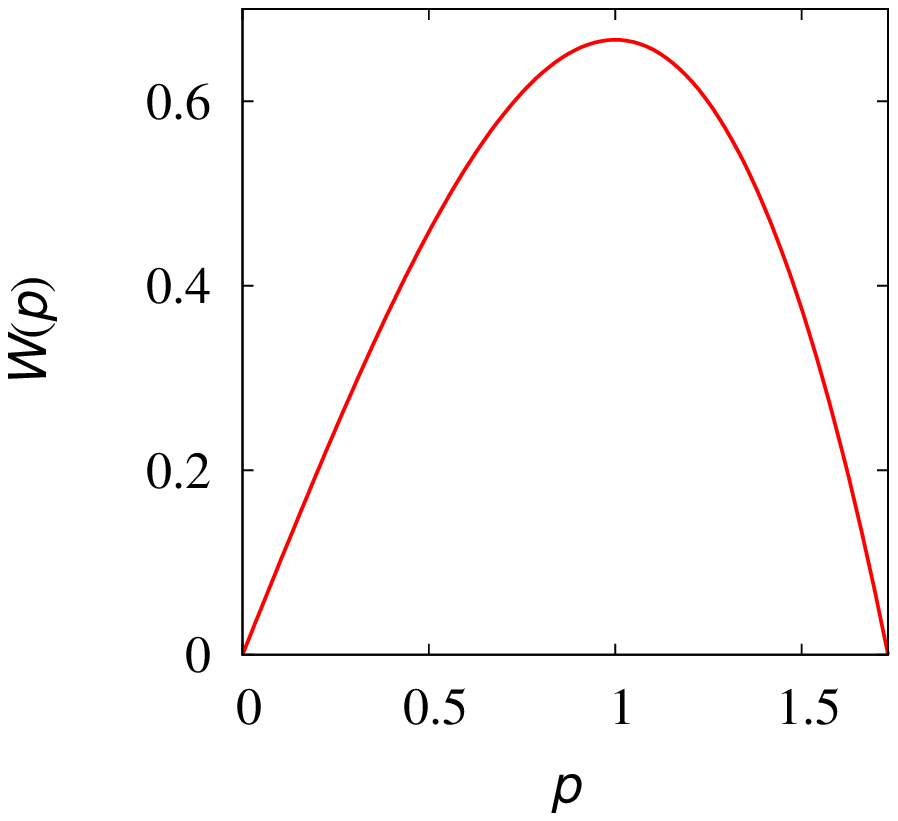} \hspace{-1.5cm} {\large (b)}
\end{center}
\caption{Graph of the characteristic function $W(p)$
defined by  eq. (\ref{eq4_23}) in the Kac limit for the
different model of random walk with exclusion treated
in the main text. Panel (a): curve (a) refers to TASEP;
curve (b) to the simple exclusion model. Panel (b)
refers to the fermionic transport model.}
\label{Fig3}
\end{figure}

\begin{table}
\begin{center}
\begin{tabular}{cc}
\hline
 & \\
Model & Function $W(p)$  \\
 & \\
\hline
 & \\
 TASEP with no bias & $p^2 \, (1-p)$ \\
& \\
Simple exclusion RW & $p \, (1-p)$ \\
 & \\
Fermionic RW with exclusion & $p \, (1-p^2/3)$ \\
& \\
\hline
\end{tabular}
\end{center}
\caption{Functional form of the function $W(p)$ entering
eq. (\ref{eq4_23}) for the Kac limit of the random
walk models satisfying an exclusion principle considered
in the main text. ``RW'' stands for ``Random Walk''.}
\label{Tab2}
\end{table}
 In all these models, for physically admissible values of $p$,
the function $W(p)$ displays a non-monotonic behavior (see figure \ref{Fig3}), 
corresponding to the
 occurrence of  a  local negative  effective 
diffusivity $D_{\rm eff}(p)=d W(p)/dp <0$.
This phenomenon, that is exclusively a consequence of the assessment of
some form of exclusion dynamics, generates instabilities,
the full characterization of which is addressed in \cite{giona_numerics}
both in a thermodynamic perspective and via numerical experiments.
The study of these phenomena permits to highlight clearly the
meaning of the different hydrodynamic limits and the role
of correlations in these paradigmatic examples of simple
particle interaction.

\section{Inclusion of potentials}
\label{sec5}

The inclusion of potentials within the GPK formalism
of interacting particle systems is relatively straightforward.
The presence of a potential contribution in particle motion, 
expressed as a functional
of the concentration field,
modifies the transition probabilities \cite{latent_heat}.
The same effect occurs for the GPK model associated 
with a system of interacting particles.

In order to analyze this phenomenology consider the simplest GPK process,
namely the classical Poisson-Kac process on the real line define
by the classical Kac's equation $d x(t)= b_0 \, (-1)^{\chi(t;\lambda_0)} 
\, d t$,
where $\chi(t;\lambda_0)$ is a usual Poisson process
with transition rate $\lambda_0$,
  the statistical characterization of which involves
the two partial probability density functions $p_+(x,t)$, $p_-(x,t)$
\cite{kac,gpk1}.
 Furthermore, assume that the nominal diffusivity
is fixed and equal to $b_0^2/2 \lambda_0=D_0$.

Consider a generic potential $f$, that in the GPK formalism
can be regarded as a functional of $p_\pm(x,t)$,
eventually depending on both space $x$ and time $t$ explicitly,
\begin{equation}
f[p_+,p_-;x,t,b_0]= C(b_0) \, F[p_+,p_-;x,t,b_0]
\label{eq5_1}
\end{equation}
For reasons that it will be soon clear, we assume that the
potential
depends explictly on the basic parameter of the Poisson-Kac
process, i.e., on $b_0$ as indicated in the functional
dependence (\ref{eq5_1}), as $\lambda_0$ is constrained by
the actual value of the diffusivity $D_0$.
In eq. (\ref{eq5_1}) we have added a prefactor $C(b_0)$,
apparently in a redundant way as $F$ depends on $b_0$,
for reasons that are related to the assessment of
the hydrodynamic limit as developed below.

In the presence of a potential, the transition
probabilities are no longer constants and equal to each other, but
are explicit functions of the potential. Specifically,
\begin{eqnarray}
\mbox{Prob}[\chi(t)=1 \, | \, \chi(t^-)=-1] &\sim & (1+f) \nonumber \\
\mbox{Prob}[\chi(t)=-1 \, | \, \chi(t^-)=1] &\sim & (1-f)
\label{eq5_2}
\end{eqnarray}
where $t^-=\lim_{\varepsilon \rightarrow 0} t-\varepsilon$, $\varepsilon>0$.
The presence of a potential exerts its action exclusively on the functional
form of the transition probability matrix, that becomes,
 through $f[p_+,p_;x,t,b_0]$,
 a functional of the partial probability densities,
\begin{equation}
{\bf A}[p_+,p_-] =
\left (
\begin{array}{cc}
\frac{1+f[p_+,p_-]}{2} & \frac{1+f[p_+,p_-]}{2} \\
\frac{1-f[p_+,p_-]}{2} & \frac{1-f[p_+,p_-]}{2}
\end{array}
\right )
\label{eq5_3}
\end{equation}
where, for short, $f[p_+,p_-]= f[p_+,p_-;x,t,b_0]$, and $f \in [-1,1]$. 
Correspondingly, the original Poisson-Kac process
in the presence of potentials admits the GPK representation
\begin{equation}
d x(t) = b_{\chi_2(t;\lambda_0 {\bf 1}; {\bf A}[p_+,p_-])} \, dt
\label{eq5_4}
\end{equation}
corresponding to a $2$-state nonlinear GPK dynamics
characterized by ${\bf b}=(b_0,-b_0)$,  ${\boldsymbol \Lambda}= \lambda_0 \, 
(1,1)$, and by the transition probability matrix expressed by eq. 
(\ref{eq5_3}). The statistical description
of eq. (\ref{eq5_4}) involves the two partial densities
$p_\pm(x,t)$ satisfying the nonlinear evlution
equations
\begin{eqnarray}
\frac{\partial p_+(x,t)}{\partial t} & = & - b_0 \, \frac{\partial 
p_+(x,t)}{\partial x} - \frac{\lambda_0}{2} \, \left [
(1- f[p_+,p_-]) \, p_+(x,t)
- (1+f[p_+,p_-]) \, p_-(x,t)  \right ]\nonumber \\
 \frac{\partial p_-(x,t)}{\partial t} & = &  b_0 \, \frac{\partial 
p_-(x,t)}{\partial x} + \frac{\lambda_0}{2} \, \left [
(1- f[p_+,p_-]) \, p_+(x,t)
- (1+f[p_+,p_-]) \, p_-(x,t) \right ] \nonumber \\
\label{eq5_5}
\end{eqnarray}
The analysis of the Kac limit of this process is particularly interesting as it
reveals  novel features of this hydrodynamic limit
in the presence of potential interactions and it disclosures the eventual
occurrence  of new physical phenomena  associated with
the emergence of phase-transitions, and field-bifurcation phenomena.
Letting $p=p_++p_-$, the balance equation for the overall concentration
coincides with eq. (\ref{eq3_26}),
 where $J_p(x,t)=b_0 \, [p_+(x,t)-p_-(x,t)]$ is a solution of
the equation
\begin{eqnarray}
\frac{\partial J_p}{\partial t} & = & - b_0^2 \, \frac{\partial p}{\partial x}
- \lambda_0 \, b_0 \, \left [ (1-f) \, p_+ - (1-f) \, p_- \right ] 
\nonumber \\ 
& =  & - b_0^2 \, \frac{\partial p}{\partial x} -\lambda_0 \, b_0 \, \left [
\frac{J_p}{b_0} - \, f \, p \right ]
\nonumber \\
& = &  - b_0^2 \, \frac{\partial p}{\partial x} -\lambda_0 
\, J_p + \lambda_0 \, b_0 \, f \, p
\label{eq5_6}
\end{eqnarray}

In order to obtain a proper Kac limit in the
presence of potential interactions it is not sufficient to
consider $b_0, \lambda_0 \rightarrow \infty$, 
keeping fixed the nominal diffusivity
$D_0$, as the quantity $b_0 \, f$ enters in the
constitutive equation for the flux and its asymptotic
 properties
are essential in the assessment of the limit.
As regards this quantity, set
\begin{equation}
b_0 \, f[p_+,p_-;x,t,b_0]= b_0 \, C(b_0) \,
F[p_+,p_-;x,t;b_0]
\label{eq5_7}
\end{equation}
 substitute for $p^\pm$ their expressions in terms of $p$ and $J_p$,
and consider the limit
\begin{equation}
\lim_{b_0 \rightarrow \infty} b_0 \, C(b_0) \,
F \left [ \frac{1}{2} \left ( p + \frac{J_p}{b_0} \right ),
\frac{1}{2} \left ( p - \frac{J_p}{b_0} \right ),x,t,b_0 \right ]
= F^*[p,J_p,x,t]
\label{eq5_8}
\end{equation}
Assume that $C(b_0)$ is given by a physical model and its
functional dependence on $b_0$ is  fixed.
In this case it can be always assumed $C=1$, since the
functional dependence on $b_0$ is contained in the
functional form of $F$.

 Three situations can occur,
as regards the limit functional $F^*[p,J_p,x,t]$:
\begin{enumerate}
\item the limit $F^*$ given by eq. (\ref{eq5_8}) exists and defines
a smooth non trivial functional $F^*$ of $p$ and $J_p$;
\item the limit $F^*$ is uniformly vanishing and consequently
the effect of the potential  is negligible in the Kac limit;
\item the limit (\ref{eq5_8}) is diverging at some point,
$F^*$ does not exists, and the Kac limit of the process
cannot be defined.
\end{enumerate}
The latter case is further addressed subsequently. 
To begin with, consider   the  first possibility, which
is obviously the most interesting one for physical reasons.
Eq. (\ref{eq5_6}) can be rewritten as
\begin{equation}
\frac{1}{\lambda_0} \, \frac{\partial J_p}{\partial t}=
- 2 \, D_0 \, \frac{\partial p}{\partial x} -J_p +b_0 \, f \, p
\label{eq5_9}
\end{equation}
that, in the Kac limit, becomes
\begin{equation}
J_p =  F^*[p,J_p;x,t] \, p - 2 \, D_0 \, \frac{\partial p}{\partial x}
\label{eq5_10}
\end{equation}
that formally corresponds to a constitutive 
equation for the flux expressed by the superposition
of a ``convective flux $F^* \, p$ and a diffusive
flux $-2 \, D_0 \,\partial p/\partial x$.

The latter interpretation is correct if  and only if 
$F^*$ depends solely on
the overall concentration $p$, and not on $J_p$, i.e.,
$F^*[p;x,t]$. Substituting eq. (\ref{eq5_10}) in this case into the
balance equation (\ref{eq3_26}),
 a nonlinear advection-diffusion equation for $p$
is obtained
\begin{equation}
\frac{\partial p(x,t)}{\partial t} = - \frac{\partial }{\partial x} \left
[ F^*[p;x,t] \, p(x,t) \right ] + 2 \, D_0 \, \frac{\partial^2 p(x,t)}{\partial x^2}
\label{eq5_11}
\end{equation}
This equation can display non-local features, depending on the
nature of the functional $F^*[p;x,t]$ that in general
may depend on the whole spatial concentration profile, and not only
on the local value $p(x,t)$ at $(x,t)$.

But there is another case, namely that $F^*[p,J_p;x,t]$
would depend explicitly on the flux $J_p$. The structure of the implicit
flux constitutive equation opens up a wealth of potentially
interesting physical and mathematical issues, depending
whether  eq. (\ref{eq5_10}) can be explicited or not, and on the
nature of the resulting explicit expression of the flux $J_p$
in terms of $p$ and $-\partial p/\partial x$.
Below, we discuss qualitatively some typical cases,
leaving a thorough investigation of this subject to
forthcoming works.

To begin with consider the case where eq. (\ref{eq5_10}) can
be explicited with respect to $J_p$, i.e.,
there exists a functional $K[p,-2 \, D_0 \, \partial p/\partial x;x,t]$,
such that
\begin{equation}
J_p(x,t)= K  \left [ p, - 2 \, D_0 \, \frac{\partial p}{\partial x};x,t 
\right ]
\label{eq5_12}
\end{equation}
fulfilling the functional equation
\begin{equation}
K \left [ p, - 2 \, D_0 \, \frac{\partial p}{\partial x}; x,t \right ]
= F^* \left [ p, K \left [ p, - 2 \, D_0 \, \frac{\partial p}{\partial x} ; x,t \right ];x,t \right ] - 2 \, D_0 \,\frac{\partial p(x,t)}{\partial  x}
\label{eq5_13}
\end{equation}
so that the resulting  balance equation for $p(x,t)$ becomes 
\begin{equation}
\frac{\partial p(x,t)}{\partial t}= - \frac{\partial }{\partial x}
\left \{ K  \left [ p, - 2 \, D_0 \, \frac{\partial p}{\partial x} ;x,t \right ]
\right \}
\label{eq5_14}
\end{equation}
It is easy to observe that  the constitutive equation (\ref{eq5_14})
is no longer Fickian, i.e., the probability flux  is no longer,
in general, 
 proportional to the probability gradient. Moreover, for  a suitable choice
of the functional $f$, it may occur locally, i.e., at some $x$ and $t$
that 
\begin{equation}
J_p(x,t) \, \frac{\partial p(x,t)}{\partial x} \geq 0
\label{eq5_15}
\end{equation}
i.e., that the flux $J_p$ is oriented towards the
direction of increasing probability gradients (uphill diffusion).
This situation is analyzed, via some numerical examples,
in \cite{giona_numerics}.

The other situation occurs in the case eq. (\ref{eq5_10}) cannot
be explicited with respect to $J_p$, leading to
a multivalued expression for the flux as a function of the
concentration gradient. In this case, multiple branches
of the flux-concentration gradient constitutive equation
can occur, determining new physical phenomena.
In order to provide a first qualitative understanding of this
class of problem, assume for simplicity that the
functional $F^*$ reduces to a local function solely of $J_p(x,t)$,
i.e.,$F^*[J_p,p;x,t]=f^*(J_p(x,t))$.

The functional form of $f^*(J_p)$ cannot be completely arbitrary,
as the original functional $f$ admits a probability interpretation
and consequently is should be bounded by $|f| \leq 1$.
This prevents, for example, the physical occurrence of a global
relation of the form $f^*(J_p)= c \, J_p^2$, where $c$ is a 
constant, as it would imply for sufficiently large $b_0$
\begin{equation}
f[p_+,p_-;x,t] \simeq \frac{c \, J_p^2}{b_0} = c \,  b_0  (p_+-p_-)^2
\label{eq5_16}
\end{equation}
that  attains arbitrarily large values for large $b_0$
and generic $(p_+-p_-)$.

\begin{figure}[!]
\begin{center}
\epsfxsize=10.cm
\epsffile{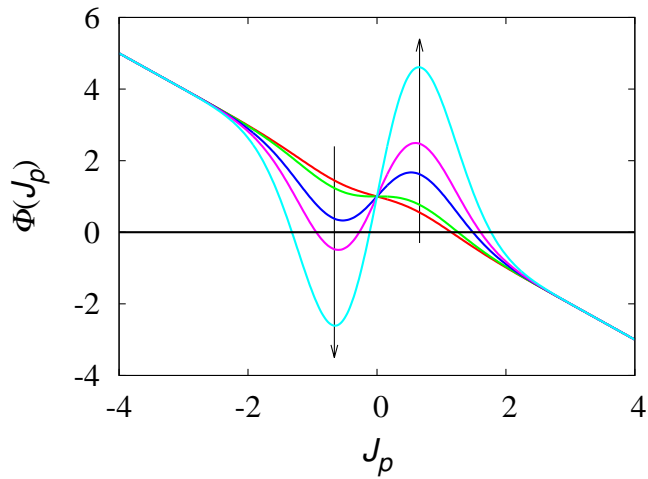}
\end{center}
\caption{Function $\Phi(J_p;p,z)$ vs $J_p$ defined by eq. (\ref{eq5_19})
 for different
values of $p$ at $z=1$. The arrows indicates increasing
values of $p=0.5,\,1,\,3,\,5,\,10$.}
\label{Fig4}
\end{figure}

It is therefore reasonable, from the above
observation, that in a physical model the  function
$f^*(J_p)$ should diverge as most linearly with $J_p$, i.e.,
\begin{equation}
f^*(J_p)= J_p \, h(J_p)
\label{eq5_17}
\end{equation}
where $h(J_p)$ is a bounded function of $J_p$, such that
$|h(J_p) \, (p_+-p_-)| \leq 1$.
For instance  a model of the form
\begin{equation}
f^*(J_p)= c \, J_p \, e^{-\beta \, J_p^2}
\label{eq5_18}
\end{equation}
where $c$ and $\beta$ are positive constants, satisfies this condition.
Setting $z= - 2 \, D_0 \, \frac{\partial p}{\partial x}$, the
 flux constitutive equation attains in this case the
expression
\begin{equation}
\Phi(J_p;p,z) = -J_p + c \, p \, J_p \, e^{-\beta \, J_p^2} + z =0
\label{eq5_19}
\end{equation}
Figure  \ref{Fig4} depicts the behavior of the function $\Phi(J_p;p,z)$
vs $J_p$ at $z=1$ for different
 values of $p$ (henceforth, we set $c=\beta=1$ a.u.), showing
that there exists a critical value  $p_c$ of $p$ above which 
the constitutive equation displays three difference branches, associated
with the solution of eq. (\ref{eq5_19}).

\begin{figure}[!]
\begin{center}
\epsfxsize=10.cm
\epsffile{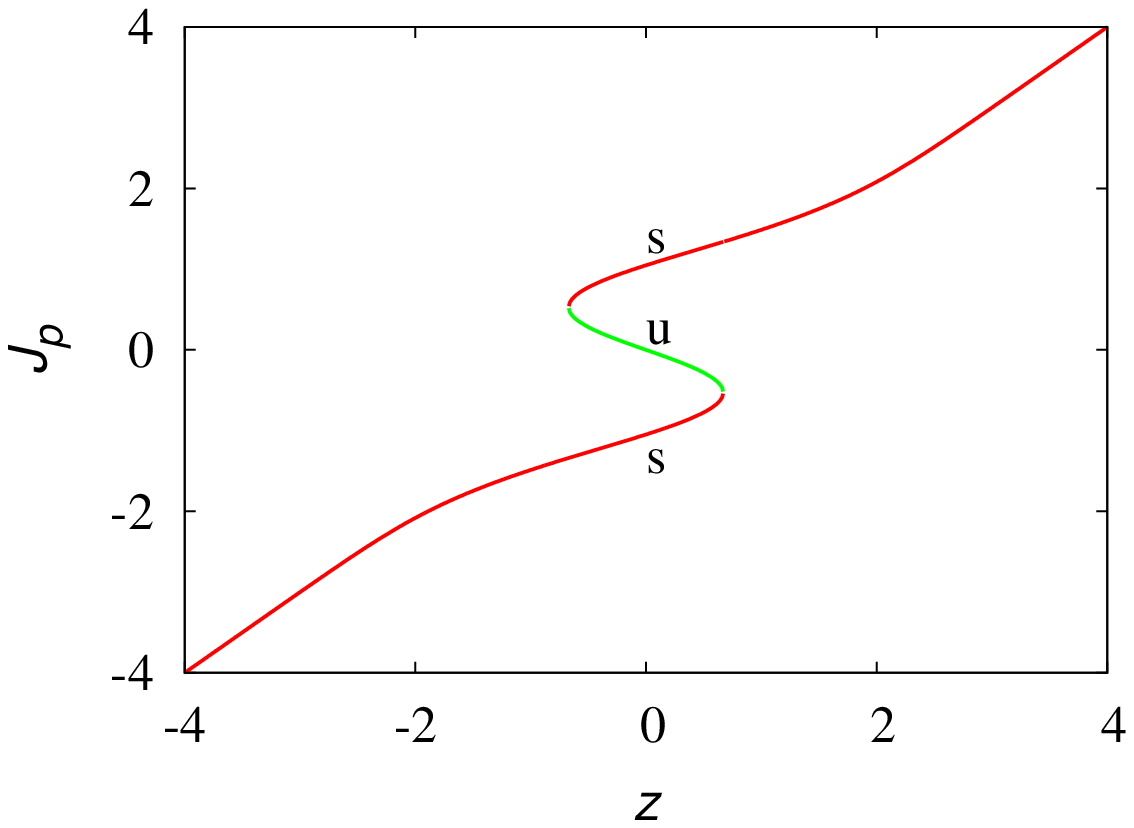} \hspace{-1.0cm} {\large (a)} \\
\epsfxsize=10.cm
\epsffile{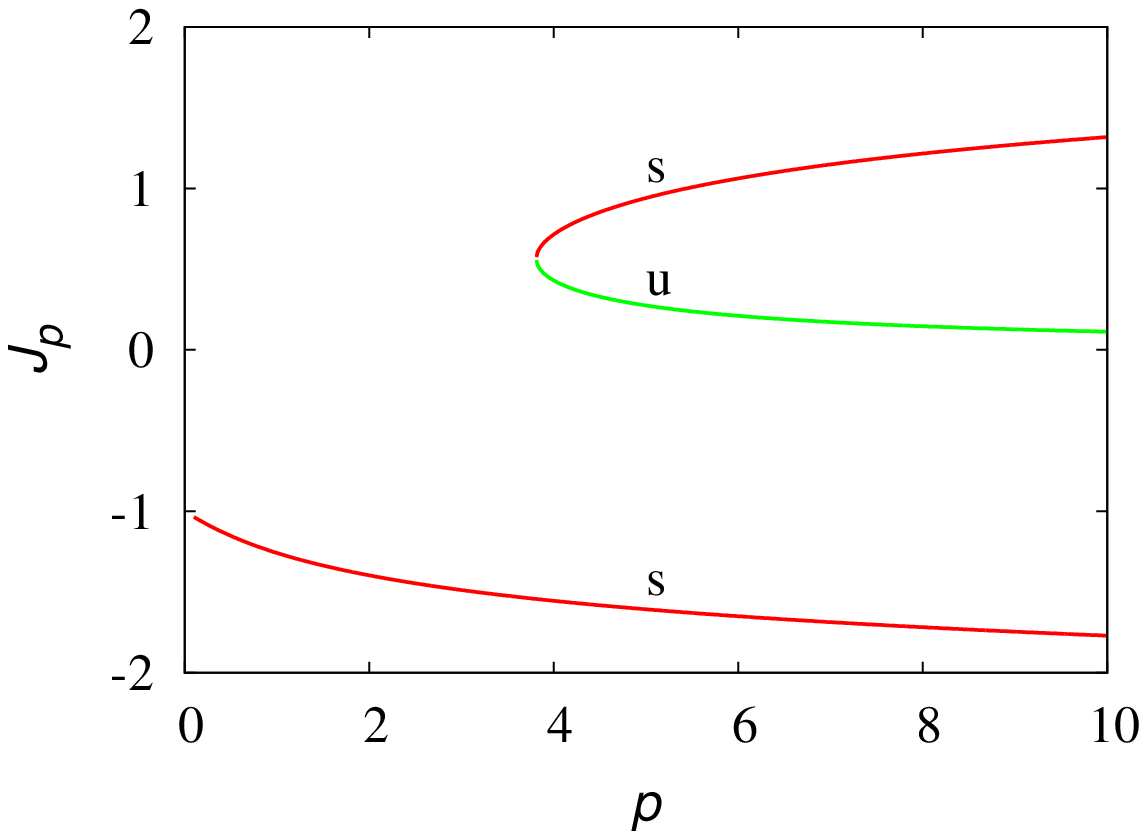} \hspace{-1.0cm} {\large (b)} 
\end{center}
\caption{Panel (a) Constitutive branches of the flux/concentration 
gradient constitutive equation ($J_p$ vs $z$) for the
model system eqs. (\ref{eq5_17})-(\ref{eq5_19}) at $p=3$.
Panel (b) $J_p$ vs $p$ at $z=-1$ for the same problem.
The labels "$s$" and "$u$" indicate respectively the stable and unstable 
branches.}
\label{Fig5}
\end{figure}

The analysis of this problem is essentially a classical
bifurcation problem of equilibria. Figure \ref{Fig5}
panel (a) depicts the constitutive equation, i.e.,
the graph of $J_p$ vs $z=-2 \, D_0 \, \partial p/\partial x$
at $p=3$, indicating the occurrence of two saddle-node
bifurcations, generating the transition from a sigle to
a three-fold structure. The stability of
the constitutive branches, follows directly from
the observation that for large $b_0$,
\begin{equation}
\frac{1}{\lambda_0} \, \frac{\partial J_p}{\partial t} \simeq \Phi(J_p,p,z)
\label{eq5_20}
\end{equation}
indicating that a constitutive branch is stable provied that
$\partial \Phi(J_p;p,z)/\partial J_p <0$, and unstable in the
opposite case.

A similar bifurcation diagram depicting $J_p$ vs $p$ at a fixed value of
$z$ is shown in panel (b) of figure \ref{Fig5}.

It follows from  the graphs depicted in figure \ref{Fig5} (a) that,
along the stable constitutive branches, the flux $J_p$
is a monotonically increasing function of $z$, as expected from
thermodynamic consistency, i.e., $\partial J_p/\partial z >0$,
and the oppositive holds for the unstable branches.

\begin{figure}[!]
\begin{center}
\epsfxsize=10.cm
\epsffile{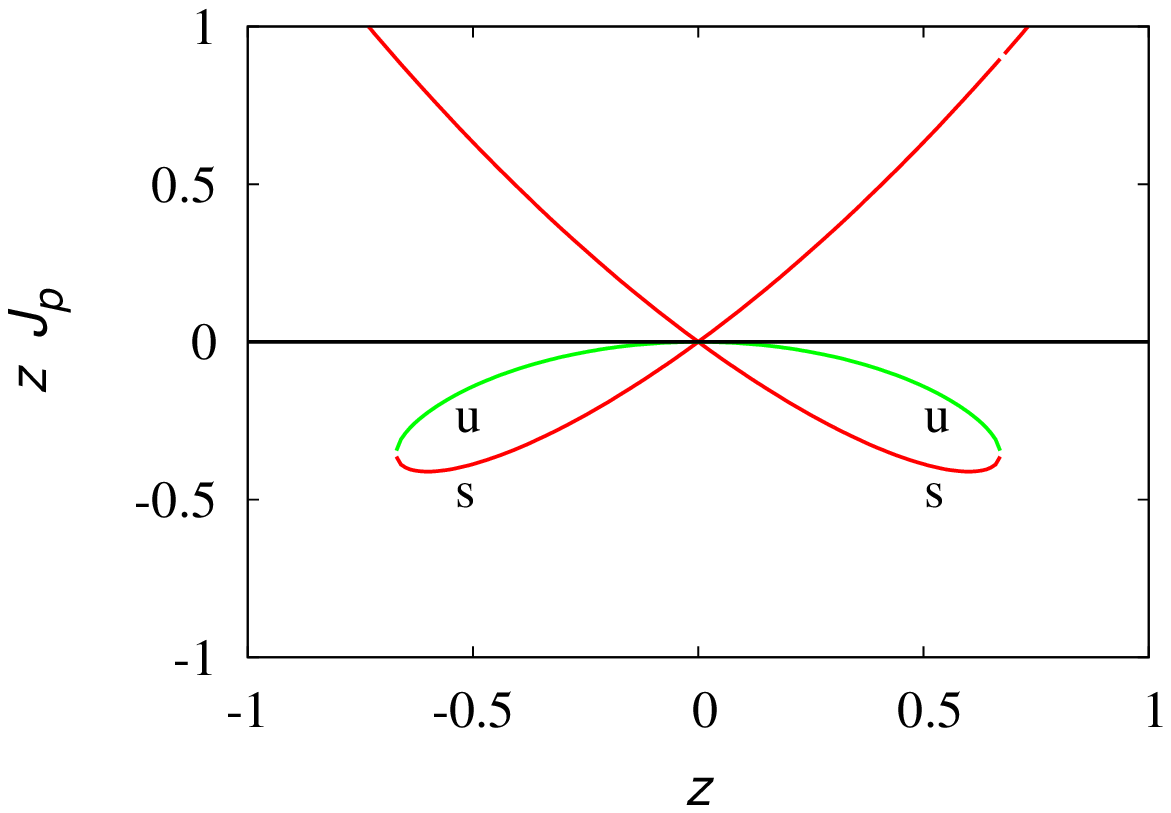}
\end{center}
\caption{Function  $z \, J_p(z;p)$ vs $z$, where
$J_p(z;p)$ are the solution of eq.  (\ref{eq5_19})
for the data depicted in figure \ref{Fig5} panel (a).}
\label{Fig6}
\end{figure}

Nevertheless, it can be observed that, for small absolute values
of $z$, the product $J_p \, z$ can be negative, even for the
stable constitutive branches, indicating the possibility
of local uphill diffusion phenomena. This
result is clearly shown in the graph depicted in figure \ref{Fig6}.

To sum up, it has been shown via a very simple example, that
system of interacting particles may give rise, in the Kac limit,
to implicit constitutive equations, producing multiple
constitutive branches. The bifurcations associated
with the explicit representation of the constitutive
equation for the flux as a function of the concentration gradient
may give rise to new classes of non-equilibrium phase transitions,
the physical and mathematical characterization of which is 
still to be developed, and it will be approached in forthcoming
works.

\section{Concluding remarks}
\label{sec6}

This article has introduced the formal setting of hyperbolic
transport models for systems of interacting particles
by considering either lattice dynamics subjected to
 simple exclusion principles or the presence of interaction potentials.
The latter phenomenology involves, in a hyperbolic continuous setting,
solely the functional dependence of the transition probability
matrix on the partial probability density functions.

For further applications, the hyperbolic formalism in the presence
of interaction potentials deserves   particular attention, as the
resulting hydrodynamic models may display a wealth of non trivial
dynamic phenomena. These phenomena are intrinsically associated with
the hyperbolic nature of the model implying bifurcations and multiplicity
of flux/concentration-gradient constitutive equations.
This is particularly evindent in the Kac limit of these
models, where, depending on the interaction potentials,
a multiplicity of constitutive equations may appear. 
The phenomelogy associated with these dynamic instabilities is 
analysed in \cite{giona_numerics}.

\end{document}